\documentclass[a4paper,UKenglish,cleveref, autoref,thm-restate,nolineno]{socg-lipics-v2021}

\pdfoutput=1 
\hideLIPIcs  

\usepackage[utf8]{inputenc}
\usepackage[T1]{fontenc}
\usepackage{algorithm}
\usepackage{algpseudocode}

\usepackage{xspace}
\usepackage{graphicx}
\usepackage{tikz}
\usepackage{multirow}
\usepackage{enumerate}
\usepackage{hyperref}
\hypersetup{
  pdftitle = {},
  pdfauthor=  {},
  colorlinks = true,
  linkcolor = black!30!red,
  citecolor = black!30!green
}
\usepackage{amsmath, amsthm, amssymb}
\newtheorem*{utheorem}{Theorem}
\newtheorem*{ulemma}{Lemma}
\newtheorem*{ucorollary}{Corollary}

\usepackage{thmtools}
\usepackage{thm-restate}
\usepackage{todonotes}

\makeatletter
\newcommand\RedeclareMathOperator{%
  \@ifstar{\def\rmo@s{m}\rmo@redeclare}{\def\rmo@s{o}\rmo@redeclare}%
}
\newcommand\rmo@redeclare[2]{%
  \begingroup \escapechar\m@ne\xdef\@gtempa{{\string#1}}\endgroup
  \expandafter\@ifundefined\@gtempa
     {\@latex@error{\noexpand#1undefined}\@ehc}%
     \relax
  \expandafter\rmo@declmathop\rmo@s{#1}{#2}}
\newcommand\rmo@declmathop[3]{%
  \DeclareRobustCommand{#2}{\qopname\newmcodes@#1{#3}}%
}
\@onlypreamble\RedeclareMathOperator
\makeatother

\DeclareMathOperator{\conf}{conf}

\newcommand{\cC}{\mathcal{C}}

\newcommand{\G}{\mathcal{G}}
\newcommand{\C}{\mathcal{C}} 
\newcommand{\F}{\mathcal{F}} 
 
\newcommand{\I}{\mathcal{I}}

\DeclareMathOperator\grid{\boxplus}

\DeclareMathOperator\tw{\textsf{tw}}
\DeclareMathOperator\td{\textsf{td}}
\DeclareMathOperator\bd{\textsf{bd}}
\DeclareMathOperator\vc{\textsf{vc}}

\newcommand{\disto}[1]{\operatorname{\textsf{dist-to-}#1}}
\newcommand{\vcpb}{\textsc{Vertex Cover}\xspace}

\newcommand{\VC}{VC\xspace}

\newcommand{\ruleref}[1]{(\hyperref[#1]{R\ref*{#1}})}

\renewcommand{\O}{\mathcal{O}}

\newcommand{\NPH}{NP-hard\xspace}
\newcommand{\ETH}{ETH\xspace}
\newcommand{\SQGM}{SQGM\xspace}
\newcommand{\ASQGM}{ASQGM\xspace}

\def\cqedsymbol{\ifmmode$\lrcorner$\else{\unskip\nobreak\hfil
\penalty50\hskip1em\null\nobreak\hfil$\lrcorner$
\parfillskip=0pt\finalhyphendemerits=0\endgraf}\fi}

\newcommand{\paramprobl}[4]{
  \begin{flushleft}
    \fbox{
      \begin{minipage}{0.97\linewidth}
        \noindent {\sc #1}\\
        {\bf Instance:} #2\\
        {\bf Question:} #3\\
        {\bf Parameter:} #4
      \end{minipage}}
    \medskip
  \end{flushleft}
}

\graphicspath{{figs/}}

\title{Subexponential algorithms for structural parameterizations of Vertex Cover in geometric graphs}

\titlerunning{Subexponential algorithms for structural parameterizations}

\author{Marin Bougeret}{LIRMM, Université de Montpellier, CNRS, Montpellier, France.}{marin.bougeret@lirmm.fr}{https://orcid.org/0000-0002-9910-4656}{}
\author{Juan Pablo Bravo}{LIRMM, Université de Montpellier, CNRS, Montpellier, France.}{juan-pablo.bravo-garrido@lirmm.fr}{https://orcid.org/0009-0008-4731-9445}{}
\author{Daniel Gonçalves}{LIRMM, Université de Montpellier, CNRS, Montpellier, France.}{daniel.goncalves@lirmm.fr}{https://orcid.org/0000-0003-3228-9622}{}

\authorrunning{M. Bougeret, J.P. Bravo, D. Gonçalves}

\Copyright{}



\relatedversion{}

\EventEditors{John Q. Open and Joan R. Access}
\EventNoEds{2}
\EventLongTitle{42nd Conference on Very Important Topics (CVIT 2016)}
\EventShortTitle{CVIT 2016}
\EventAcronym{CVIT}
\EventYear{2016}
\EventDate{December 24--27, 2016}
\EventLocation{Little Whinging, United Kingdom}
\EventLogo{}
\SeriesVolume{42}
\ArticleNo{23}

\begin{document}

\maketitle

\begin{abstract}
A number of recent works have investigated the existence of parameterized subexponential algorithms in geometric graph classes
for hitting a fixed small graph as a subgraph or as a minor. In this setting, given an $n$-vertex graph, the goal is to decide in time
$n^{\O(1)}2^{o(k)}$ if a solution of size at most $k$ exists.
Inspired by recent results in kernelization for structural parameters,
we investigate subexponential algorithms for the Vertex Cover (VC) problem, but parameterized by distance-to-triviality instead of solution size.
More formally, given a fixed "trivial" graph class $\C$, the $VC/\disto{\C}$ problem takes as input $(G,X,k)$, where $G \setminus X \in \C$, and
asks if $G$ admits a vertex cover of size at most $k$. Our goal is to design algorithms for this problem running in time $n^{\O(1)}2^{o(|X|)}$.
The motivation is that, since $|X|$ is typically arbitrarily smaller than $k$, such algorithms might be faster than ETH-tight subexponential algorithms parameterized by $k$.

We identify two sets of sufficient conditions under which $VC/\disto{\C}$ admits a subexponential algorithm: one for the case where $\C$ is the class of paths, and another for the case where $\C$ is the class of
graphs of constant treedepth. We also provide two applications of these algorithmic results: we prove that $s$-string graphs and pseudo-disk graphs satisfy the first set of conditions, whereas string graphs satisfy the second set. The two sets share a common condition that might be of independent interest, as it implies a property
that bounds the treewidth of any string graph in terms of a given modulator to a fixed graph class $\C$.

Finally, concerning more restricted graph classes having the $ASQGM(\omega)$ property (such as map graphs and $H$-minor free graphs),
we prove the following dichotomy: for any class $\G$ with the $ASQGM(\omega)$ property and any minor-closed class $\C$,
$VC/\disto{\C}$ restricted to input graphs in $\G$ admits a subexponential algorithm if and only if $\C$ has constant treewidth.
\end{abstract} 


\section{Introduction}

\subsection{Context and motivation.}
\paragraph*{Combining subexponential algorithms and structural parameterizations.\\}
In this paper, we study the \vcpb (\VC) problem from the perspective of parameterized complexity, with the goal of obtaining faster algorithms under refined parameterizations.

In the classical parameterization (where the parameter is the solution size), given an instance $(G,k)$, it is known under the Exponential Time Hypothesis (\ETH) that there is no algorithm for \VC running in time $n^{\mathcal{O}(1)}2^{o(k)}$, and even no $2^{o(n)}$-time algorithm, on general graphs~\cite{Cygan2015Book}.

There are two main directions to circumvent these lower bounds. The first is to seek fixed-parameter tractable (FPT) algorithms with respect to smaller parameters, that is, algorithms running in time $f(p(G)) \cdot n^{\mathcal{O}(1)}$, where the parameter $p(G)$ can be significantly smaller than $k$ (for instance, $p(G)$ could be the treewidth of $G$). The second direction is to design subexponential parameterized algorithms, that is, algorithms running in time $n^{\mathcal{O}(1)}2^{o(k)}$, which is only possible on restricted graph classes under \ETH.

The first direction fits within the classical scope of parameterized complexity: given a graph parameter $p$, determine the complexity of \VC parameterized by $p$ (denoted $\VC/p$), in particular whether it is FPT, W[1]-hard, or admits XP algorithms. This line of work has been extensively studied, including for \vcpb, for instance in the context of parameterizations above or below structural guarantees~\cite{kellerhals2022vertex}.

The second direction has also seen significant developments. Bidimensionality theory provides subexponential algorithms of the form $n^{\mathcal{O}(1)}2^{o(k)}$ (typically $n^{\mathcal{O}(1)}2^{\mathcal{O}(\sqrt{k})}$) for a wide range of problems, including \vcpb and \textsc{Feedback Vertex Set} for example, on planar graphs and more generally on bounded genus and $H$-minor-free graphs~\cite{demaine2005subexponential}. More recently, these techniques have been extended to geometric intersection graphs which offer a natural way to go beyond sparse graph classes while still retaining forms of underlying geometric structure. In this line of work we can mention \cite{bonnet2019optimality} that studies the existence of $2^{o(n)}$ algorithm for many problems in string graphs, \cite{marx2016subexponential} studying subexponential parameterized algorithms for graphs of polynomial growth, and \cite{korhonen2023induced} that provides $2^{o(n)}$ algorithm for many problems (including \VC) in induced-minor free graphs, implying an $n^{\O(1)}2^{o(k)}$ algorithm by observing that the textbook linear kernel for \VC via crown reduction outputs an induced subgraph, which will remain induced-minor free.
Notice that string graphs are a special case, since they exclude the 1-subdivision of $K_5$ as an induced minor.

This naturally suggests combining these two approaches, and studying subexponential algorithms under structural parameterizations in geometric graph classes. Let us recall that the combination of such parameters and geometric graphs is somehow necessary, as even when parameterizing by $k$, one needs to restrict to particular graph classes. For instance, it was asked in~\cite{lokshtanov2025subexponential} whether subexponential algorithms for $\mathcal{F}$-hitting problems (which generalize \vcpb) could be obtained when parameterizing by $\tw(G)$ (denoting treewidth) instead of $k$.

\paragraph*{Focusing on distance-to-triviality parameters.\\}
We argue that distance-to-triviality parameters provide a particularly promising framework for this line of research. In this type of measure, introduced in~\cite{isomorphism-elimination-distance}, one fixes a graph class $\mathcal{C}$ (such as independent sets or forests), and defines $\disto{\mathcal{C}}(G)$ as the minimum size of a vertex set $X \subseteq V(G)$ such that $G \setminus X \in \mathcal{C}$.

One motivation for studying such parameters, as opposed to classical width measures like treewidth or treedepth ($\td$),
is first that, even on planar graphs (see \autoref{thm:negativetwtd}), there is no subexponential algorithm for $\VC/\tw$, or even for $\VC/\td$. 
This suggests a parallel with kernelization: it is also relatively straightforward~\cite{bougeret2019much} to rule out polynomial kernels for $\VC/m$ for many graph parameters $m$ satisfying $m(G_1 \cup G_2) = \max(m(G_1), m(G_2))$ (such as $\tw$ or $\td$). In contrast, parameterizations of the form $m = \disto{\mathcal{C}}$ have led to a rich and well-structured theory. In this setting, one is given an instance $(G,X,k)$, where $X$ is a modulator such that $G \setminus X \in \mathcal{C}$\footnote{Providing the modulator $X$ as part of the input is standard in this setting, as it separates the task of computing a modulator from that of kernelization.}, and the goal is to compute an equivalent instance $(G',k')$ with $|V(G')| = |X|^{\mathcal{O}(1)}$. Such kernels can be significantly smaller than the classical $\mathcal{O}(k)$-vertex kernel for \vcpb, since $\disto{\mathcal{C}}(G)$ can be much smaller than $k$.

The first positive result in this direction was due to Jansen~\cite{JansenB13}, who obtained a polynomial kernel for $\VC/\disto{\text{trees}}$, corresponding to the case where $X$ is a feedback vertex set, which is a typical example where $|X|$ can be arbitrarily smaller than $k$.
This result initiated a line of work aiming to characterize for which classes $\mathcal{C}$ the problem $\VC/\disto{\mathcal{C}}$ admits a polynomial kernel. Subsequent results provide many separated positive/negative
results for specific classes $\mathcal{C}$, until a complete characterization for minor-closed classes was proved in~\cite{vertex-cover-bridge-depth}, where it was proved that 
for any minor closed $\mathcal{C}$, $\VC/\disto{\mathcal{C}}$ admits a polynomial kernel if and only if $\C$ has constant bridge depth ($\bd$). The bridge depth is a graph measure, which in particular verifies $\tw(G) \le \bd(G) \le \td(G)$ for any graph $G$. This motivates the following general question.

\medskip
\noindent
\textbf{For which geometric graph classes $\mathcal{G}$ and graph classes $\mathcal{C}$ does $\VC/\disto{\mathcal{C}}$ admit a subexponential parameterized algorithm?}

\paragraph*{Our results.\\}

Toward addressing this question, we provide several initial results, which we group into two categories depending on the considered graph class $\mathcal{G}$.
We point out that our positive results are stated in an abstract way: rather than relying on ad hoc geometric arguments for each class, we identify a set of properties under which our algorithms apply. We then show separately that these conditions are satisfied by well-studied geometric graph classes. 
This approach yields algorithms that are robust, in the sense that they do not require an explicit geometric representation of the input graph.

 \begin{figure}[!ht]
 \centering           
     \includegraphics[scale=1]{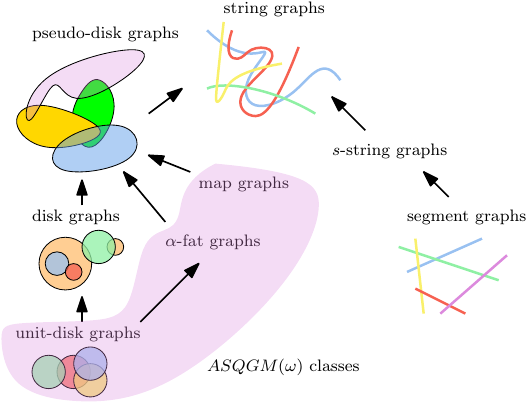}
     \caption{Inclusion between graph classes.}
     \label{fig:graphclasses}
 \end{figure}

We first consider graph classes that satisfy the $\ASQGM(\omega)$\footnote{$\ASQGM$ means Almost SubQuadratic Grid Minor, by opposition to $\SQGM$ property, where there exists $\epsilon >0$ such that $\tw(G)=\O(\grid(G)^{2-\epsilon})$.} property~\cite{berthe24ASQGM}, that is, classes in which there exists $\epsilon > 0$ such that the treewidth can be bounded as $\tw(G) = \mathcal{O}(\omega(G)^{\mathcal{O}(1)} \cdot \grid(G)^{2-\epsilon})$, where $\grid(G)$ denotes the largest integer $C$ such that $G$ contains a $C \times C$ grid as a minor, and $\omega(G)$ denotes the size of a maximum clique. This condition already captures several classical classes (see \autoref{fig:graphclasses}) such as planar graphs, map graphs, $H$-minor-free graphs, and unit disk graphs~\cite{berthe24ASQGM}. For these classes, we obtain the following dichotomy for $\C$, which holds both for $\mathcal{F}$-hitting and minor-$\mathcal{F}$-hitting. Given an instance $(G,X,k)$, the objective in these problems is to decide whether there exists a set $S \subseteq V(G)$ of size at most $k$ such that $G \setminus S$ contains no graph from $\mathcal{F}$ as a subgraph (or as a minor, respectively).

\begin{utheorem}
    Let $\mathcal{G}$ be a hereditary graph class that has the $\ASQGM(\omega)$ property, and let $\mathcal{C}$ be a hereditary graph class having bounded treewidth.
    Let $\F$ be a finite set of connected graphs  such that (minor)-hitting$_{\F}$ can be solved in $n^{\O(1)}2^{\O(\tw(G)\log(\tw(G)))}$.
    
    Then, there exists an $\epsilon > 0$ such that, when restricted to input graphs in $\mathcal{G}$, (minor)-hitting$_{\F}$ can be solved in time $n^{\O(1)}2^{|X|^{1-\epsilon}}$.
\end{utheorem}

For a large range of families $\F$ and $\G$, the following observation enlightens a sharp dichotomy depending on  $\tw(\mathcal{C})$.
\begin{observation}
    \label{obs:lbASQGM}
    Let $\F$ be a finite set of connected graphs on at least two vertices, and such that at least one of them is planar.
    Let $\mathcal{C}$ a minor-closed graph class that has unbounded $\tw(\C)$, and let $\G$ be the class of planar graphs.
  Then, unless $P=NP$, (minor)-hitting$_{\F}$|$\disto{\mathcal{C}}$ in $\G$ is not in $XP$.
\end{observation}
Indeed, a minor-closed family $\mathcal{C}$ of unbounded treewidth contains all planar graphs, and this implies that
$\disto{\mathcal{C}}(G)=0$ for every $G\in \G$.
Hence the observation follows from the fact that (minor)-hitting$_{\F}$ remains \NPH when restricted to planar graphs for any $\F$ containing connected graphs of at least two vertices and at least one planar one~\cite{yannakakis1978node}.

We note that the assumption that $\mathcal{C}$ is minor-closed also appears in dichotomy results for polynomial kernelization~\cite{vertex-cover-bridge-depth}, and that extending such results beyond minor-closed classes is typically challenging.
While this dichotomy is essentially tight for the considered setting, the underlying techniques rely on branching on large cliques, which crucially decreases the parameter $|X|$ (and not only $k$). As a consequence, this approach does not extend to graph classes such as disk graphs or 2-DIR graphs (a subclass of segment graphs corresponding to intersections of axis-parallel segments). These graphs do not belong to $\ASQGM(\omega)$ as they can contain large $K_{t,t}$ as minor, hence having $\Omega(t)$ treewidth, while having $\omega(G)=\O(1)$ (see Figure 1 in \cite{Berthe2025}).

The presence of such “hidden” $K_{t,t}$ structures, and more generally the difficulty of handling intersections of thin geometric objects, makes the problem significantly more challenging. This phenomenon already appears in the context of \textsc{Feedback Vertex Set}, where substantial effort was required to extend subexponential algorithms parameterized by $k$ from unit disk graphs to disk graphs~\cite{lokSODA22}, and further to thin classes such as 2-DIR or, more generally, $s$-string graphs~\cite{Berthe2025} (where each pair of strings intersects at most $s$ times).

Mimicking some of the first results obtained in the world of kernelization with $\VC/$ $\disto{Forest}$ and $\VC/\disto{(td=\O(1))}$ (meaning parameterizing by distance to constant treedepth graphs),
we next establish results on more general graph classes, which constitute the main contributions of this paper.
To state them, we need to introduce several technical conditions.
The first condition concerns the behavior of treewidth. It is motivated by \autoref{thm:tw-string-XR} showing that string graphs satisfy this property, which may be of independent interest.

\begin{definition}\label{def:degreeTW}
Let $\epsilon_r,\epsilon_E,\epsilon_\Delta > 0$. A graph class $\G$ has \emph{$(\epsilon_r,\epsilon_E,\epsilon_\Delta)$-modulator-bounded treewidth} if every $G\in \G$ verifies the following. For every partition $V(G) = R \cup X$ and every $\Delta \le |X|^{\epsilon_r}$ such that $\Delta(G[X])\le \Delta$, and  $|N_X(C)| \le \Delta$ for every connected component $C$ of $G[R]$, we have
\[
\tw(G) = \mathcal{O}\big(\tw(G[R]) + |E(G[X])|^{\epsilon_E} + |X|^{1-\epsilon_\Delta}\Delta^{\epsilon_\Delta}\big).
\]
\end{definition}

Notice that this property does not hold for general graphs. Indeed, if we take $G$ as a subdivided $K_n$, where $X$ corresponds to the $n$ vertices of the original clique, and each of the $\frac{n(n-1)}{2}$ connected component of $R$ corresponds to a degree two vertex.
In this case, $\tw(G)=n-1$, and as $\tw(G[R])=1$, $|E(G[X])|=0$, and $\Delta=2$, there is no eligible $\epsilon_\Delta > 0$.

Let us now explain why this property is crucial to obtain positive results. Recall that in our setting, we are given an instance $(G,X,k)$ of $\VC/$ $\disto{\C}$ (where $\C$ is typically a class with constant treewidth), where $G[R] \in \C$. As explained more in details in \autoref{sec:techniques}, our goal is to reduce to $tw(G)=o(|X|)$, implying a $n^{\O(1)}2^{o(|X|)}$ algorithm using standard dynamic programming. To that end, our algorithm performs some branching to obtain $\Delta(G[X]) \le |X|^\epsilon$, and to ensure that every connected component $C$ of $G[R]$ has $|N_X(C)|\le \Delta$ for $\Delta \le |X|^\epsilon$ (for some well chosen $\epsilon>0$). Once this is achieved, notice that the bound on treewidth in \Cref{def:degreeTW} is such that $tw(G)=o(|X|)$ as desired.

The second condition concerns neighborhood complexity introduced in~\cite{reidl2019characterising}. In its original form, this notion counts the number of distinct neighborhoods that vertices can have within a set $X$. Here, we instead consider neighborhoods of connected subgraphs of $G \setminus X$, and formulate the condition in a parameterized way, as it is crucial to keep the dependence on $|X|$ close to linear. First, given a graph $G$ and a set $X \subseteq V(G)$, let us denote $d_X(v) = |N(v) \cap X|$ and $\Delta_X(G) = \max_{v \in V(G)}{d_X(v)}$. Second, observe that $\Delta(G[X]) = \max_{v \in X}{d_X(v)}$ can be less than $\Delta_X$. Furthermore, given a value $d$, let
  \begin{eqnarray*}
    \mathcal{C}^{\le d}(G,X) & = & \{C \mid C \text{ is a connected component of } G\setminus X \text{ such that } |N_X(C)| \le d\} \text{ and}\\
    \mathcal{C}^{> d}(G,X) & = & \{C \mid C \text{ is a connected component of } G\setminus X \text{ such that } |N_X(C)| > d\}.
  \end{eqnarray*}

\begin{definition}\label{def:pboundedNC}
  Let $p$ be a polynomial. A graph $G$ has \emph{$p$-bounded neighborhood complexity} if for any $X \subseteq V(G)$, and any $d$, then
\[
\big|\{N_X(C) \mid C \in \mathcal{C}^{\le d}(G,X)\}\big| \le p(\Delta(G[X]),d)|X|.
\]
We say that a graph class $\G$ has $p$-bounded neighborhood complexity if for any $G \in \G$, $G$ has $p$-bounded neighborhood complexity. 
\end{definition}

The final condition is motivated by our study of $\VC/\disto{(td=\O(1))}$. In this setting, each connected component $C$ of $G \setminus X$ has constant treedepth. Intuitively, components $C$ with $|N_X(C)| \le p_1(t)$ (for a polynomial $p_1$ and $t=|X|^\epsilon$) are "easy" to handle: if all components were of this type, then \autoref{def:degreeTW} would already imply $\tw(G) = o(|X|)$. Therefore, it is necessary to control the number of components with large neighborhoods.

\begin{definition}\label{def:boundedLNC}
Let $p_1,p_2$ be polynomials. A graph $G$ has \emph{$(p_1,p_2)$-bounded large neighborhood components} with respect to some function $g$, if for every vertex partition $V(G)=R \cup X$, we have either $|X| = \O(\Delta_X(G)^5)$, or
\[
|\mathcal{C}^{>p_1(\Delta_X(G))}(G,X)| \le p_2(\Delta_X(G)) \cdot g(\td(G[R])) \cdot |X|.
\]
We say that a graph class $\G$ has $(p_1,p_2)$-bounded large neighborhood components if for any $G \in \G$, $G$ has $(p_1,p_2)$-bounded large neighborhood components with respect to the same function $g_{\G}$.
\end{definition}

We are now ready to define our two generic classes.

\begin{definition}\label{def:Type1Class}
A hereditary graph class $\mathcal{G}$ is a \emph{path forest modulator compliant} class (a {\em PATH-Mod compliant} class for short) with parameters $(p,\epsilon_r,\epsilon_E,\epsilon_\Delta)$ if $\G$ has $p$-bounded neighborhood complexity and $\G$ has $(\epsilon_r,\epsilon_E,\epsilon_\Delta)$-modulator-bounded treewidth.
\end{definition}

\begin{definition}\label{def:Type2Class}
A hereditary graph class $\mathcal{G}$ is a \emph{bounded treedepth modulator compliant} class (a {\em TD-Mod compliant} class for short) with parameters $(p_1,p_2,\epsilon_r,\epsilon_E,\epsilon_\Delta)$ if $\G$ has $(p_1,p_2)$-bounded large neighborhood components and $\G$ has $(\epsilon_r,\epsilon_E,\epsilon_\Delta)$-modulator-bounded treewidth.
\end{definition}

\begin{utheorem}[see \autoref{thm:treedepth} for a full statement]
$\VC/\disto{(td=\O(1))}$ admits a subexponential algorithm on every TD-Mod compliant class.
\end{utheorem}

\begin{utheorem}[see \autoref{thm:path} for a full statement]
$\VC/\disto{\text{paths}}$ admits a subexponential algorithm on every PATH-Mod compliant class.
\end{utheorem}

To apply these theorems to well known classes, we also prove that known classes satisfy the above conditions;
these structural results may be of independent interest.

\begin{theorem}\label{thm:ss-pd-type12}
    The classes of $s$-string graphs and pseudo-disk graphs are both
    PATH-Mod compliant, whereas the class of string graphs is TD-Mod compliant.
\end{theorem}

As a corollary, we obtain our main algorithmic consequence.

\begin{ucorollary}
The problem $\VC/\disto{\text{paths}}$ admits subexponential algorithms on $s$-string graphs and pseudo-disk graphs, whereas $\VC/\disto{(td=\O(1))}$ admits a subexponential algorithm on string graphs. 
\end{ucorollary}

Finally, observe that in the setting of $\VC/\disto{(td=\O(1))}$, we have $\td(G) \le |X| + \mathcal{O}(1)$, since a treedepth decomposition of $G$ can be obtained by placing all vertices of $X$ above the decomposition of $G \setminus X$. Therefore, a subexponential algorithm for $\VC/\td$ would strictly improve our result; however, such an algorithm is ruled out under \ETH (see \autoref{thm:negativetwtd}).

\subsection{Organisation of the paper}
In \autoref{sec:techniques} we present our techniques, explaining in particular what are the challenges specific to structural parameterization, and how we address them. 
\autoref{sec:prelim} contains technical preliminaries.
We present first positive results for $\VC/\disto{(td=\O(1))}$ (\autoref{sec:bounded-treedepth})
and $\VC/\disto{\text{paths}}$ (\autoref{sec:paths}), for TD-Mod compliant classes and PATH-Mod compliant classes, respectively.
The easiest case of $\ASQGM$ classes is handled in
\autoref{sec:asqgm}.
Then, \autoref{sec:graph} is devoted to graph theoretical results where we prove that $s$-string and pseudo-disk graphs are indeed PATH-Mod compliant classes, whereas string graphs are a TD-Mod compliant class.
Finally, negative results when parameterizing by $\tw$ or $\td$ instead of $\disto{\C}$ is presented in \autoref{sec:negative}, and we conclude in \autoref{sec:conclusion}.
\section{Algorithmic techniques}
\label{sec:techniques}
We first introduce a common ingredients of our algorithms, the branchings that reduce our inputs into relatively sparse ones (in \autoref{sec:techniquebranching}), and different ways to bound the treewidth of intersection graphs (in \autoref{sec:techniquetreewidth}). Then we will sketch our algorithm for $\VC/\disto{(td=\O(1))}$ (in \autoref{sec:sketchtd}),
and the one for $\VC/\disto{\text{paths}}$ (in \autoref{subsec:techniquesPath}).

A very common scheme to obtain subexponential algorithms parameterized by $k$ for hitting problems such as $\VC$ is to
branch on dense structures, and once this is no longer possible, show that the absence of such structures implies that $tw(G)=o(k)$, and then solve the problem in time $2^{\tw}$ using dynamic programming.
Let us now discuss these steps and explain the challenges when parameterizing by $\VC/\disto{\mathcal{C}}$ instead of $k$.

\subsection{Branching on dense structures.}\label{sec:techniquebranching}
Consider a measure $p(G,k)$, where $0 \le p(G,k) \le p_0$, and a branching procedure that stops branching when the measure reaches a negative value. 
Given a fixed $\epsilon > 0$, we further assume that for an instance $(G,k)$ there is at most one branch where $p$ decreases by $1$, and $p_0^{\mathcal{O}(1)}$ branches where $p$ decreases by $p_0^{\epsilon}$.
Now, let $f(p,n)$ denote the worst-case running time of an algorithm performing such branching, for any input $(G,k)$ where $p(G,k) \le p$ and $|V(G)| \le n$. If each branching takes polynomial time, we get
\[f(p,n) \le n^{\mathcal{O}(1)} + f(p-1,n) + p_0^{\mathcal{O}(1)} f(p-p_0^{\epsilon},n).\] 
It is easy to prove that this implies $f(p,n) \le pn^{\mathcal{O}(1)} + p_0^{\mathcal{O}(1)} f(p-p_0^{\epsilon},n)$, and hence
\[f(p,n) = \mathcal{O}\big(n^{\mathcal{O}(1)} p_0^{\mathcal{O}(p_0^{1-\epsilon})}\big),\]
which is subexponential in $p_0$.
Typically, when the measure $p(G,k)$ is the solution size $k$, it is folklore for $\VC$ to branch on vertices $v$ with $d(v) \ge k^{\epsilon}$ (by taking either $v$ or $N(v)$ in the vertex cover), or on bicliques or cliques of size $c = k^{\epsilon}$.

When considering an instance $(G,X,k)$ of $\VC/\disto{\mathcal{C}}$ with a measure corresponding to the parameter, that is $p(G,X,k)=|X|$, notice that:
\begin{itemize}
\item branching on a vertex $v \in X$ with $d_X(v) \ge |X|^{\epsilon}$ is allowed (meaning that we can reduce to $\Delta(G[X]) \le |X|^{\epsilon}$ in subexponential time in $X$),
\item branching on a vertex $v \notin X$ with $d_X(v) \ge |X|^{\epsilon}$ is not allowed, as it leads to
$f(p,n) \le n^{\mathcal{O}(1)} + f(p,n-1) + f(p-|X|^{\epsilon},n)$,
which resolves to $f(p,n)=\Omega(n^{|X|^{1-\epsilon}})$ (meaning that we cannot reduce to $\Delta_X(G) \le |X|^{\epsilon}$ in subexponential time in $X$, unless $n$ is polynomial in $|X|$),
\item branching on a vertex $v \in X$ with $d_R(v) \ge |X|^{\epsilon}$ (where $R = V(G) \setminus X$) is not allowed as there is no branch where $|X|$ decreases by $|X|^{\epsilon}$.
\end{itemize}
This is a first price to pay when parameterizing by $|X|$ instead of $k$, and we will see in the next paragraphs how we handle this for $\VC/\disto{\text{paths}}$ and $\VC/\disto{(td=\O(1))}$.

\subsection{Bounding the treewidth}
\label{sec:techniquetreewidth}
Let us start with two important known bounds on treewidth.
Recall that in any class in $ASQGM(\omega)$ (such as unit disk graphs), there exists $\lambda,\epsilon > 0$ such that
$\tw(G) = \mathcal{O}\big(\omega(G)^\lambda \cdot \grid(G)^{2-\epsilon}\big)$.

These classes are the easiest to handle; let us now sketch how to handle them.
When considering $\VC/k$, we can obtain $\omega(G) = \mathcal{O}(k^{\epsilon})$ by branching on large cliques, and $\grid(G) \le \sqrt{k}$ as $\VC$ is a bidimensional problem (if $\grid(G) > \sqrt{k}$ then the answer is no). This leads to $\tw(G) = o(k)$.

When considering $\VC/\disto{\mathcal{C}}$ where $\tw(\mathcal{C}) \le c$ for some constant $c$, we observe that given an input $(G,X,k)$, any large clique $K$ satisfies $|K \cap X| \ge |K| - c$, and thus $K$ lies almost entirely inside $X$. Therefore, branching on $K$ also decreases $|X|$ by $\Omega(|X|^{\epsilon})$.
Thus, in time subexponential in $|X|$, we can reduce to the case where $\omega(G) = \mathcal{O}(|X|^{\epsilon})$.
Moreover, we also have $\grid(G) \le \sqrt{|X|c}$, leading to $\tw(G) = o(|X|)$.

For more challenging classes that are not in $ASQGM(\omega)$ (such as disk graphs, pseudo-disk graphs, or $s$-string graphs), we need different arguments.
Observe first that string graphs are in $ASQGM(cross)$ \cite{grigoriev2014bidimensionality}, where $cross$ denotes the maximum number of times a string is intersected (i.e., $\tw(G) = \mathcal{O}(cross(G)^\lambda \cdot \grid(G)^{2-\epsilon})$),
but it is not easy to reduce to $cross(G) = |X|^{\epsilon}$, suggesting that different techniques are needed to bound the treewidth.
As the classes we target are typically subclasses of string graphs, the following result of \cite{lee2016separators} is of great interest.
\begin{utheorem}[\cite{lee2016separators}]
Any $K_{t,t}$-free string graph with $n$ vertices has $\tw(G) = \mathcal{O}(\sqrt{t \log t \cdot n})$.
\end{utheorem}

We point out that this result was recently extended to induced-minor-free graphs in \cite{korhonen2023induced}.
This theorem is very useful for $\VC/k$, since branching on vertices of degree at least $t = k^{\epsilon}$ implies that the graph becomes $K_{t,t}$-free.
Then one can apply the classical linear kernel based on crown reductions, obtaining an induced subgraph $(G',k')$, which remains a string graph and satisfies $\tw(G') = o(k)$.

This technique might also be usable for $\VC/\disto{\mathcal{C}}$, but even if we manage to branch to obtain $K_{t,t}$-free graphs for $t = |X|^{\epsilon}$, we would still need a linear kernel in $|X|$ (or $|X|^{1+\epsilon}$ would be sufficient) that preserves the graph class.
This is challenging: for example, the kernel for $\VC/\disto{\text{forest}}$ reduces path lengths using local operations that do not preserve string graphs.
Moreover, known kernels have size $\mathcal{O}(|X|^3)$ for $\VC/\disto{\text{forest}}$~\cite{JansenB13}, and $\O(|X|^{2^{c^2}})$ for 
$\VC/\disto{(td\le c)}$~\cite{bougeret2019much}, and it has been proved in \cite{hols2022elimination} that unless $NP \subseteq coNP/Poly$, then there is no polynomial kernel with size $|X|^{2^{c}-\epsilon}$.
Even if we allow here subexponential time, it seems to remain difficult.
Instead, we prove the following result for string graphs, which we use both for $\VC/\disto{\text{paths}}$ and $\VC/\disto{(td=\O(1))}$.

\begin{utheorem}
Consider a string graph $G$, a partition of its vertices into two sets $X,R$, and a value $\Delta \le |X|^{1/4}$. If $\Delta(G[X])\le \Delta$, and if $|N_X(C)| \le \Delta$ for every connected component $C$ of $G[R]$, then,
$\tw(G) = \mathcal{O}\big(m_X^{1/2} + \tw(G[R]) + |X|^{5/6}\Delta^{1/6}\big)$.
\end{utheorem}

Observe that this corresponds to having $(\tfrac{1}{4},\tfrac{1}{2},\tfrac{1}{6})$-modulator bounded treewidth.
Given an instance $(G,X,k)$ of $\VC/\disto{\text{paths}}$ or
$\VC/\disto{(td=\O(1))}$ and a value $t$, recall that $\mathcal{C}^{> t}(G,X)$ is the set of connected components $C$ of
$G \setminus X$ such that $|N_X(C)| > t$.
Recall that we can branch to ensure that $d_X(v) \le |X|^{\epsilon}$ for any $v \in X$, and that in our setting $R = V(G) \setminus X$ corresponds to paths or graphs of constant treedepth (and thus constant treewidth).
Therefore, if $\mathcal{C}^{> t}(G,X) = \emptyset$ for $t = |X|^{\epsilon}$, then the above theorem implies that
$\tw(G) = \mathcal{O}(|X|^{\frac{5+\epsilon}{6}})$.

In other words, it suffices to handle the components in $\mathcal{C}^{> t}(G,X)$, which is done differently for $\VC/\disto{\text{paths}}$ and $\VC/\disto{(td=\O(1))}$, as we explain next.

\subsection{Sketch of proof for \texorpdfstring{$\VC/\disto{(td=\O(1))}$}{VC/dist-to-(td=\O(1))}}\label{sec:sketchtd}
As discussed above, we can reduce to the case where $\Delta(G[X]) \le t$ for $t=|X|^{\epsilon}$, and the next goal is to also obtain
$\Delta_X(G) \le t$.

To that end, we use the fact that the polynomial kernel for
$\VC/\disto{(td=\O(1))}$ of \cite{bougeret2019much} outputs an induced subgraph. Hence, given $(G,X,k)$ (with $\td(G \setminus X) \le c$) 
and $G \in \mathcal{G}$, for any hereditary graph class $\G$, we can first apply this kernel to obtain an equivalent instance $(G',X',k')$ such that $G' \in \mathcal{G}$ and $|V(G')| = |X|^{\mathcal{O}_c(1)}$.
Then, branching on vertices $v \in G' \setminus X'$ with $d_{X'}(v) \ge |X'|^{\epsilon}$ becomes subexponential in $|X|$.

Without loss of generality, let us now rename the obtained instance $(G,X,k)$ and observe that $\Delta_X(G) \le |X|^{\epsilon}$.
We now return to the case where $\mathcal{C}^{> t}(G,X) \neq \emptyset$. The key ingredient to handle these components is the following theorem.

\begin{utheorem}
Let $G$ be a string graph and consider any vertex partition $V(G)=X \cup R$.
Either $|X| = \O(\Delta_X(G)^5)$, or 
$|\mathcal{C}^{\ge 6\td \Delta_X^2}(G,X)| = \O_{\td}(\Delta_X(G)^7) \cdot |X|$, where $\td(G\setminus X) \le \td$.
\end{utheorem}

Observe that this corresponds to the property of $(\O(\Delta_X(G)^2),\O(\Delta_X(G)^7))$-bounded large neighborhoods.
Applying this theorem with $\Delta_X(G) = t$,
one obtains $|\mathcal{C}^{> \O(t^2)}(G,X)| = \mathcal{O}_c(|X|^{1+\mathcal{O}(\epsilon)})$.
Note also that this theorem cannot be used for $\VC/\disto{\text{paths}}$ as 
treedepth of path is arbitrarily large, which explains why we proceed differently for $\VC/\disto{\text{paths}}$.
We can now use a recursive argument (inspired by  \cite{bougeret2019much}) as follows.
Let $Z$ be the set of roots of treedepth decompositions of the components in $\mathcal{C}^{> \O(t^2)}(G,X)$, which implies that $\td(C) \le c-1$, 
for any connected component $C\in \mathcal{C}^{> \O(t^2)}(G,X\cup Z)$. 
(Note that the components of $\mathcal{C}^{\le \O(t^2)}(G,X\cup Z)$ may have treedepth greater than $c-1$.)
We then recurse on $(G', X \cup Z, k')$.
As the size of the modulator only slightly increases since 
$|X \cup Z| = \mathcal{O}(|Z|) = \mathcal{O}_c(|X|^{1+\mathcal{O}(\epsilon)})$,
and as the recursion depth is $c$, a constant, this leads to a subexponential algorithm in $|X|$. 

\subsection{Sketch of proof for \texorpdfstring{$\VC/\disto{\text{paths}}$}{VC/dist-to-paths}}
\label{subsec:techniquesPath}
As before, we start by branching to ensure that $\Delta(G[X]) \le t$ where $t=|X|^{\epsilon}$.
The next goal is to obtain the same bound for vertices $v \notin X$.
Unlike the bounded-treedepth case, we cannot rely on the only existing polynomial kernel~\cite{JansenB13}, as it requires local merging operations that do not preserve being a string graph for example.
Instead, we rely on neighborhood complexity, and prove the following result.

\begin{utheorem}
There exists a polynomial $p$ such that for any graph $G$, an $s$-string graph (for a constant $s$) or a pseudo-disk graph, any set $X \subseteq V(G)$, and any $d$, then
\[
\big|\{N_X(C) \mid C \in \mathcal{C}^{\le d}(G,X)\}\big| \le p(\Delta(G[X]),d)|X|.\]
\end{utheorem}

This corresponds to having $p$-bounded neighborhood complexity.
Now, given an instance $(G,X,k)$ of $\VC/\disto{\text{paths}}$ (with $R = V(G) \setminus X$), since paths are bipartite, we can partition $R$ into two independent sets $\mathcal{Z}_1, \mathcal{Z}_2$.
Applying the theorem to each graph $G_i=G[X \cup \mathcal{Z}_i]$ (with $d = |X|$) yields
$NC(G,X) =_{def} |\{N_X(v) \mid v \in R\}| = |X|^{\mathcal{O}(1)}$.
Thus, if there exists a vertex $v \in R$ with $N_X(v) = X'$ and $|X'| \ge t$, instead of branching on $v$, we consider the set
$Y = \{v' \in R \mid N_X(v') = X'\}$,
and branch by either taking $X'$ or taking $Y$.
In the second branch, $NC(G,X)$ decreases by $1$, and since $NC(G,X) = |X|^{\mathcal{O}(1)}$, this yields a subexponential branching in $|X|$, and we can now assume that $\Delta_X(G) \le t$.

\begin{figure}[htbp]
    \centering
    \includegraphics[width=\linewidth]{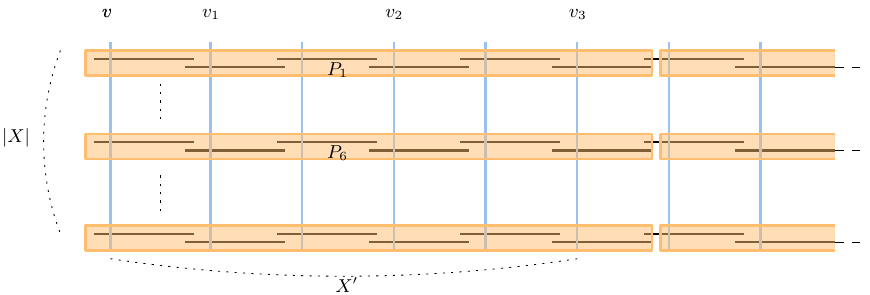}
    \caption{A 2-DIR representation of an input $(G,X,k)$ of $\VC/\disto{\text{paths}}$, where vertices of $X$ are depicted in blue. Here, even if $\Delta_X(G) \le 1$, observe that $\tw(G) \ge |X|$ as it contains a $K_{|X|,|X|}$ as a minor. The subpaths computed by $\text{cut}_t(G,X)$ (for $t=6$) are represented as orange rectangles. Here, $|\text{cut}_6(G,X)|$ contains $6$ subpaths $P_i$ and there is a subset $X'\subset X$ such that $3$ of them are such that $N_X(P_i)=X'$. Here, it implies the existence of a $K_{1,3}$ of conflict $3$, which is $(v,X''=\{v_1,v_2,v_3\})$. In particular, notice that for any $v_j$, $\conf_R(\{v,v_j\}) \ge 3$, as not taking $v$ and $v_j$ in a solutions forces to take one more vertex (compared to a minimum vertex cover) in these $3$ paths.} 
    \label{fig:sketchpath}
\end{figure}

We now return to the case where $\mathcal{C}^{> t}(G,X) \neq \emptyset$.
Unlike the treedepth case, the bound given by the theorem on bounded large neighborhoods, does not lead to $|\mathcal{C}^{> t}(G,X)| = \O(|X|^{1+\O(\epsilon)})$ as $\td(G\setminus X)$ is unbounded. Moreover, as illustrated in \autoref{fig:sketchpath}, $\tw(G)$ may still be in $\Omega(|X|)$.
To handle this we define the following cut procedure that partitions each path in $\mathcal{C}^{> t}(G,X)$ into subpaths $P_i$ by collecting vertices until $d_X(P_i) \ge t$.
Since $d_X(v) \le t$ for all $v \in R$, we can ensure that $d_X(P_i) \le 2t$, although $|V(P_i)|$ may still be arbitrarily large.
Let $\text{cut}_t(G,X)$ be the set of paths $P_i$ obtained by the cut procedure.
The key argument is a win-win analysis depending on $|\text{cut}_t(G,X)|$. 

If $|\text{cut}_t(G,X)| \le 2t \cdot p(t,2t+2)\cdot |X|$ (where $p$ is the polynomial from the $p$-bounded neighborhood complexity property), then we can prove that $\tw(G)$ is small.
Indeed, let $Z$ be the set containing the first vertex of each subpath $P_i$, define $X' = X \cup Z$, and let $P'_i = P_i \setminus Z$.
Using the modulator bounded treewidth result with $\Delta = 2t+2$ (since $d_{X'}(P'_i) \le d_X(P_i)+2 \le 2t+2$), we obtain $\tw(G) = o(|X|)$. 

Otherwise, suppose that $|\text{cut}_t(G,X)| > 2t \cdot p(t,2t+2)\cdot |X|$.
Partition $R$ into two sets $R_1,R_2$ such that every path $P_i\in \text{cut}_t(G,X)$ fully belongs to one of them, but does not belong to its adjacent paths (if any).
Using $p$-bounded neighborhood complexity on the two graphs 
$G[X\cup R_1]$ and $G[X\cup R_2]$ with $d=2t+2$, we obtain
$|\{N_X(P_i) \ |\ P_i\in \text{cut}_t(G,X) \}| \le 2 \cdot p(t,2t+2)\cdot |X|$.
Hence, there exists a set $X' \subseteq X$ and at least $t$ distinct subpaths $P_i$ such that $N_X(P_i) = X'$.
We hope now for a situation where, either we take a lot of vertices in $X'$ (which makes the parameter $|X|$ decreases as $|X'| \ge t$), or, not taking them
implies that we have to take a lot of vertices in a "suboptimal way" in the $P_i's$, which is captured by the following definitions.

\begin{definition}[Conflict]\label{def:conflict}
Let $(G,X,k)$ be an instance of $\VC/\disto{\mathcal{C}}$.
For $X' \subseteq X$ and $R' \subseteq R$, define $\conf_{R'}(X')$ as the difference of sizes between a vertex cover of $G[R']$ if the vertices of $N_{R'}(X')$ were forced into it (e.g. by the fact that we consider vertex covers of $G$ avoiding $X'$ completely), and an optimal vertex cover of $G[R']$.
Denoting $\vc(H)$ the minimum size of a vertex cover of $H$ we have that,
\[ \conf_{R'}(X') = |N_{R'}(X')| + \vc(G[R'] \setminus N_{R'}(X')) - \vc(G[R']).\]
\end{definition}
The notion of conflict is crucial in many kernelization results, as it enables win--win arguments of the following form: either a vertex $v \in X$ satisfies $\conf_R(v) > |X|$, in which case one can show that we can always take $v$ in the solution, or otherwise $G\setminus X$ should be small.
In our setting, the situation is more delicate: we want to consider vertex $v$ such that $\conf_R(v) > |X|^{\epsilon}$. For such vertices, it is no longer true that we can always take them, and therefore we need to branch.
Now, in the branch where we do not take them, we need to keep track of the fact that we lose $|X|^{\epsilon}$ compared to the local optimum on $R$.
This motivates the following definition.

\begin{definition}[Extra budget]\label{def:extra_budget}
We define $b(G,X,k) = k - \vc(G \setminus X)$, and call it the \emph{extra budget}.
\end{definition}

Observe that we may assume that $0 \le b(G,X,k) < |X|$.
Indeed, if $b(G,X,k) < 0$, then we can immediately answer no, since the budget $k$ does not even suffice to cover the edges in $R$.
On the other hand, if $b(G,X,k) \ge |X|$, then we can answer yes by taking all vertices of $X$ together with an optimal solution in $R$.
The key point relating conflict and budget is the following lemma, which shows that avoiding a subset $X'$ with large conflict significantly decreases the budget.

\begin{ulemma}
Let $(G,X,k)$ be an instance of $\VC/\disto{\mathcal{C}}$ and let $R = V(G) \setminus X$. Let $X' \subseteq X$ and let $c = \conf_R(X')$.
Then $b(G \setminus N(X'), X \setminus N(X'), k - |N(X')|) \le b(G,X,k) - c$.
\end{ulemma}

We are now ready to define the structure on which we branch. The fact that the branching is performed in subexponential time is ensured by considering the extra budget as the measure (see \Cref{sec:techniquebranching}).

\begin{definition}\label{def:K1T}
Given an instance $(G,X,k)$ of $\VC/\disto{\mathcal{C}}$ and integers $t$ and $c$, we say that
\emph{$(v,X')$ is a $K_{1,t}$ of conflict $c$ in $(G,X)$} if:
\begin{itemize}
\item $\{v\} \cup X' \subseteq X$ and $|X'| = t$,
\item for every $v' \in X'$, we have $\{v,v'\} \notin E(G)$ and $\conf_R(\{v,v'\}) \ge c$, where $R = V(G) \setminus X$.
\end{itemize}
\end{definition}

Recall that we are in a situation where there exists a set $X' \subseteq X$ and $t$ distinct subpaths $P_i$ such that $N_X(P_i) = X'$.
With a more involved version of the \emph{cut} procedure\footnote{The real \emph{cut} procedure is not defined in this short version.}, hence a slightly different set cut$_t(G,X)$, this implies the existence of a $K_{1,t'}$ $(v,X'')$ of conflict $t'$ in $(G,X)$, where $t' = |X|^{\epsilon'}$ for some $\epsilon' > 0$, and $v \cup X'' \subseteq X'$. See \autoref{fig:sketchpath} for a simple illustration.

We then branch on such a structure in the following $t'+2$ branches:
we either take $v$ in the vertex cover, or we take all vertices of $X''$ in the vertex cover, or for every $v'\in X''$ we avoid the pair $\{v,v'\}$ in the vertex cover (i.e. we take $N(\{v,v'\})$ in the vertex cover), which decreases the budget by at least $t'$.

Finally, observe that the resulting algorithm 
is conceptually simple: it searches for a $K_{1,t'}$ of conflict $t'$, and if none is found, it runs the dynamic programming algorithm.
The notion of the cut is only used in the analysis, to prove that in the absence of such a structure, the treewidth is small.

\section{Preliminaries}
\label{sec:prelim}

\subsection{Basics}
In this paper logarithms are binary and all graphs are simple, loopless and undirected.
Given a graph $G$, we denote by $\vc(G)$ the minimum size of a vertex cover of $G$, and $\omega(G)$ the maximum order of a clique in $G$.
We denote by $d_G(v)$ the degree of $v \in V(G)$, or simply $d(v)$ when $G$ is clear from the context.
We denote by $\grid_k$ the $(k,k)$-grid graph. 
Given a set of graphs $\F$, we denote $n_{\F}=\max_{G \in \F}|V(G)|$.
Unless otherwise specified we use standard graph theory terminology, as in \cite{diestel2005graph} for instance.

What we call a \emph{graph parameter} here is a function $p$ mapping any (simple) graph to a natural number and that is invariant under isomorphism.
Among graph parameters, we denote by $\grid(G)$ the maximum $k$ such that $\grid_k$ is a minor of $G$, and by $\tw(G)$ the treewidth of $G$.
\begin{definition}[Treedepth]\label{def:treedepth}
    A \emph{treedepth decomposition} of a connected graph $G$ is a rooted tree $T$ such that $V(T) = V(G)$, and for every edge $\{u, v\} \in E(G)$, vertex $u$ is an ancestor of vertex $v$ in $T$, or vice versa. A treedepth decomposition of a disconnected graph is just a disjoint union of treedepth decompositions for its connected components. 
    The \emph{treedepth} of $G$, denoted by $\td(G)$, is the minimum depth (in number of vertices) of a treedepth decomposition of $G$.
\end{definition}
Let us also recall the definition of elimination distance to a fixed class introduced in~\cite{isomorphism-elimination-distance}.
Given a fixed graph class $\C$ and any graph $G$, we denote by $\disto{\C}$ the graph parameter defined by $\min\{|X| \mid X \subseteq V(G) \mbox{ s.t. } G \setminus X \in \C \}$.

\subsection{Graph classes}
\label{sec:graphCl}

\paragraph*{Geometrical classes}
 In this article,  we are mainly concerned with geometric graphs described by the intersection of objects in the Euclidean plane.
 The most general class we consider are \emph{string graphs}, which are intersection graphs of strings (a.k.a. Jordan arcs).

We also consider the subclass of {\em $s$-string graphs} where every pair of the considered strings intersect on at most $s$ points.
We also consider pseudo-disk graphs defined as follow.
A set $\mathcal{S}$ of regions of the plane homeomorphic to disks forms a system of pseudo-disks if, for any two intersecting regions $R_1, R_2 \in \mathcal{S}$, their borders $\partial R_1$ and $\partial R_2$ either cross properly exactly twice or are tangent exactly once.
We refer the reader to \autoref{fig:graphclasses} for inclusion between these classes.

\paragraph*{(A)SQGM properties}

\begin{definition}[\cite{baste2022contraction}]
Given $c < 2$, a graph class $\G$ has the \emph{subquadratic grid minor property} for $c$ (\emph{SQGM} for short), denoted $\G \in SQGM(c)$, if $\tw(G)=\O(\grid(G)^c)$ for all $G\in \G$.
We write $\G \in SQGM$ if there exists $c<2$ such that $\G \in \SQGM(c)$.
\end{definition}
It is known that every graph $G$ satisfies the inequality $\tw(G) = \O(\grid(G))^c$ for some $c<10$~\cite{chuzhoy2021towards}, but in the SQGM property it is required that $c<2$.
For instance planar graphs and more generally $H$-minor free graph~\cite{demaine2008linearity} are in $SQGM(1)$, but in general geometrical graph classes do not have this property due to large cliques, motivating the following definition.

\begin{definition}[\cite{berthe24ASQGM}]
Given a graph parameter $p$ and a real $c < 2$, a graph class $\G$ has the \emph{almost subquadratic grid minor property for $p$} (\emph{ASQGM} for short) for $p$ and $c$ if there exists a function $f$ such that $\tw(G)=\O(f(p(G))\grid(G)^c)$.
The class $\G$ has $\ASQGM(p)$ if there exists $c<2$ such that $\G$ has the \ASQGM property for $p$ and $c$. 
\end{definition}
It is known~\cite{baste2022contraction,grigoriev2014bidimensionality} that for example, $\alpha$-fat graphs and map graphs has the $\ASQGM(\omega)$ property.

\subsection{Problems}

Given a graph class $\C$ we define the following problems.
Notice that in both problems we assume that the so-called modulator $X$ is given as input. This is a standard hypothesis used in previously mentioned results on kernelization, as it separates the difficulty of computing the modulator. Moreover, in the cases we focus on here, where $\C$ correspond to paths or graphs of constant treedepth, we can approximate the modulator in polynomial time as discussed in~\cite{vertex-cover-bridge-depth}.
\paramprobl
    {$\VC/\disto{\C}$}
    {$(G,X,k)$ where $G$ is a graph, $X \subseteq V(G)$, $G \setminus X \in \mathcal{C}$, $k$ integer}
    {Decide if there exists $\vc(G) \le k$, where $\vc(G)$ denotes the minimum size of a vertex cover of $G$.}
    {$|X|$}

Given a finite set of graphs $\F$, we define the following problem.
\paramprobl
    {minor-hitting$_{\F}$|$\disto{\C}$}
    {$(G,X,k)$ where $G$ is a graph, $X \subseteq V(G)$, $G \setminus X \in \mathcal{C}$, $k$ integer}
    {Decide if there exists $S \subseteq V(G)$ such that $|S| \le k$ and $G\setminus S$ does not contain a graph $F \in \F$ as a minor.}
    {$|X|$}

The hitting$_{\F}/\disto{\C}$ problem is defined in the same way, except that we require that $G\setminus S$ does not contain a graph $F \in \F$ as a subgraph.

\subsection{Preliminary branching on degree in the modulator}
The following lemma is folklore.
\begin{lemma}\label{lemma:prelimBranching}
Let $\G$ and $\C$ be hereditary graph classes.
There is an algorithm, that, given an instance $(G,X,k)$ of $\VC/\disto{\C}$ with $G \in \G$, and an integer $t$, produces a set $\I$ of instances of $\VC/\disto{\C}$ such that
\begin{itemize}
    \item for any $(G',X',k') \in \I$: 
    \begin{itemize}
        \item $G'$ is an induced subgraph of $G$, $X' \subseteq X$, and $k' \le k$,
        \item for any $v \in  X'$, $d_{X'}(v) \le t$,
    \end{itemize}
    \item $(G,X,k)$ is a yes-instance iff there exists $(G',X',k') \in \I$ which is a yes-instance,
    \item the algorithm runs in $n^{\O(1)}|\I|$.
\end{itemize}
Moreover, $|\I|=2^{\O(\log(|X|)\frac{|X|}{t})}$.

\end{lemma}
\begin{proof}
Let us consider an integer $t$ and an input $(G,X,k)$ of $\VC/\disto{\C}$.
Let us define algorithm $A(G,X,k)$ such that, 
while there exists $v \in X$ with $d_X(v) > t$, 
$A(G,X,k)$ outputs $A(G \setminus v, X \setminus v, k-1) \cup A(G \setminus N(v), X \setminus N(v), k-|N(v)|)$ (we assume that whenever $k<0$ then $A$ returns $\emptyset$).
Denoting $f(n,x)$ the maximum size of the output of $A(G,X,k)$ for any $G$ with $|V(G)| \le n$ and $|X| \le x$, it is not difficult to see that $f(n,x) \le f(n,x-1)+f(n,x-t)$, leading to $f(n,x) \le xf(n,x-t)$, and $f(n,x) \le x^{\frac{x}{t}}$ as claimed.
Other properties of the lemma are immediate.
\end{proof}
\section{Subexponential algorithm for \texorpdfstring{$\VC/\disto{(td=\O(1))}$}{VC/dist-to-(td=\O(1))}}\label{sec:bounded-treedepth}

\subsection{Algorithm definition}

We are now ready to describe the main algorithm, divided into three parts. In Algorithm $A$, we call a known polynomial kernel from~\cite{bougeret2019much}, which produces an induced subgraph of $G$ whose size is bounded by a function of $|X|$. This will allow  algorithm $B$ to branch on to decrease $\Delta_X(G)$.
To avoid heavy specification, we add the common requirement (to algorithm $A,B,C$) that the input graph $G$ has $(p_1,p_2)$-bounded large neighborhood components. The exact specification of the algorithms rely on $p_1$ and $p_2$, on $c$ the maximum treedepth of $G\setminus X$, and on two small values $\epsilon,\epsilon'$, whose exact value do not need to be set now.

\begin{algorithm}[!h]
\caption{$A_{\epsilon,p_1,p_2,\epsilon',c}(G,X,k)$}
\begin{algorithmic}[1]
\Require $\td(G \setminus X) \le c$ 
  \State let $X_0 = X$
  \State apply the polynomial kernel of~\cite{bougeret2019much} that outputs an equivalent instance $(G',k')$ where $G'$ is an induced subgraph of $G$ and $|V(G')| \le |X_0|^{f(c)}$ \label{line:algoTDA}
  \State let $t=|X_0|^\epsilon$ and $X' = X_0 \cap V(G')$
  \State return $B_{t,c,|X_0|,p_1,p_2,\epsilon'}(G',X',k')$
\end{algorithmic}
\label{def:algoTDA}
\end{algorithm}

We now describe the second part of the procedure. The goal of Algorithm $B$ is to simplify the instance by branching on vertices outside the modulator that have many neighbors in it. This branching is possible because for any vertex $v$, any vertex cover either contains $v$ or contains its whole neighborhood.
Here, we use the threshold $t$ defined in Algorithm $A$.

\begin{algorithm}[!h]
\caption{$B_{t,c,|X_0|,p_1,p_2,\epsilon'}(G,X,k)$}
\begin{algorithmic}[1]
\Require $\td(G \setminus X) \le c$, and $|V(G)| \le |X_0|^{f(c)}$
 \If{there exists $v \notin X$ with $d_X(v) > t$}
  \State return $B_{t,c,|X_0|,p_1,p_2,\epsilon'}(G \setminus v,X,k-1) \bigvee 
  B_{t,c,|X_0|,p_1,p_2,\epsilon'}(G \setminus N(v),X \setminus N(v),k-|N(v)|)$ \label{line:algoTDB}
  \Else
  \State let $c_0 = c$
  \State return $C_{t,c,c_0,|X_0|,p_1,p_2,\epsilon'}(G,X,k)$
 \EndIf
\end{algorithmic}
\label{def:algoTDB}
\end{algorithm}

We now describe the final part of the procedure. Algorithm
$C_{t,c,c_0,|X_0|,p_1,p_2,\epsilon'}$
is based on an induction on $c$. For each recursive call, let $i = c_0 - c$. We define $t_0 = |X_0|^{\epsilon'}$, and $t_{i+1} = t_i^{\lambda_2(i+1)}|X_0|^{\epsilon'}$, where $\lambda_2$ is such that $p_2(x) = \O(x^{\lambda_2})$.

We first branch in line~\ref{line:algoTDC1} to decrease
$\Delta(G[X])$, as adding roots to $X$ may increase the degree in
$G[X]$ after a few rounds. To handle this potential increase, we allow the thresholds $t_i$ to grow as the recursion progresses.

Let $L$ be a uniform constant such that the first case in
\autoref{def:boundedLNC} guarantees $|X| \le L\Delta_X(G)^5$. After the branching in line~\ref{line:algoTDC1}, we will have $t_i \ge \Delta_X(G)$. We therefore directly solve the instance using a standard DP in line~\ref{line:algoTDCNEW} whenever $|X| \le Lt_i^5$. Otherwise, $|X| > Lt_i^5 \ge L\Delta_X(G)^5$, so the first case of \autoref{def:boundedLNC} cannot occur.

We can then use the second case of \autoref{def:boundedLNC} to bound the number of components in $\C(G,X)^{>p_1(t_i)}$. If this set is nonempty, we add to $X$ the roots of treedepth decompositions of these components and recurse in line~\ref{line:algoTDC2}, decreasing their treedepth by one, while the treedepth of the other components of $G\setminus X$ may remain as large as $c_0$. If it is empty, the algorithm falls into the final base case in line~\ref{line:algoTDC3} and solves the instance using a standard DP.

\begin{algorithm}[!h]
\caption{$C_{t,c,c_0,|X_0|,p_1,p_2,\epsilon'}(G,X,k)$}
\begin{algorithmic}[1]
\Require for any $v \notin X, d_X(v) \le t$, $\td(G \setminus X) \le c_0$, and for any connected component $C \in \C(G,X)^{>p_1(t_i)}$  $\td(C) \le c$, where $i = c_0 - c$
  \If{there exists $v \in X$ with $d_X(v) > t_i$}
  \State return $C_{t,c,c_0,|X_0|,p_1,p_2,\epsilon'}(G \setminus v,X \setminus v,k-1) \bigvee 
  C_{t,c,c_0,|X_0|,p_1,p_2,\epsilon'}(G \setminus N(v),X \setminus N(v),k-|N(v)|)$  \label{line:algoTDC1}
  \EndIf
  \If{$|X| \le L t_i^5$}
    \State Decide $(G,k)$ using a standard DP parameterized by $\tw$ (as $\tw(G)$ is small) \label{line:algoTDCNEW}
  \EndIf
\State let $Z$ be the roots of a treedepth decomposition of components of $\C(G,X)^{>p_1(t_i)}$ ($|Z|$ is small by \autoref{def:boundedLNC})
  \If{$Z \neq \emptyset$}
    \State Let $X' = X \cup Z$
    \State return $C_{t+1,c-1,c_0,|X_0|,p_1,p_2,\epsilon'}(G,X',k)$ \label{line:algoTDC2}
  \Else
  \State Decide $(G,k)$ using a standard DP parameterized by $\tw$ (as $\tw(G)$ is small) \label{line:algoTDC3}
  \EndIf
\end{algorithmic}
\label{def:algoTDC}
\end{algorithm}

Correctness of this algorithm is immediate as the only branching we perform in algorithms $A,B,C$ is to take a vertex or its neighbors, in the vertex cover. To bound its running time and decide how to set the parameters $\epsilon,\epsilon'$, we refer to the next subsection.

\subsection{Running time analysis}

\begin{theorem}\label{thm:treedepth}
Let $\G$ be a TD-Mod compliant class with parameters $(p_1,p_2,\epsilon_r,\epsilon_E,\epsilon_\Delta)$ where $p_1(x) = \O(x^{\lambda_1})$ and $p_2(x) = \O(x^{\lambda_2})$. Then, $\VC/\disto{(td=\O(1))}$ restricted to graphs in $\G$ can be solved in parameterized subexponential time. More precisely, we can solve an instance $(G,X,k)$ of $\VC/\disto{(td=\O(1))}$ in time $n^{\O(1)}2^{\log(|X|)|X|^{1-\epsilon^\star}}$, where \[\epsilon^\star = \min(\frac{1}{5M_{\td}+1},\frac{\epsilon_r}{M_{\td}\lambda_1+\epsilon_r},\frac{1-\epsilon_E}{M_{\td}(1+\lambda_2\td)\epsilon_E+1},\frac{\epsilon_{\Delta}}{M_{\td}(\lambda_1\epsilon_{\Delta}+\lambda_2\td(1-\epsilon_{\Delta}))+1})\] and $M_{\td} = \sum_{i=0}^{\td} \frac{(\td)!}{(\td - i)!}\lambda_2^i$.
\end{theorem}

\begin{proof}

Let $\epsilon' \geq \epsilon > 0$ that we will fix later. Recall that the algorithm is composed of three parts, namely Algorithms $A$, $B$, and $C$, which are executed sequentially. At a high level, the total running time can be expressed as the cost of the preprocessing step performed by Algorithm $A$, followed by the branching procedure of Algorithm $B$, and the cost of solving each resulting instance using Algorithm $C$. 

Firstly, the worst case running time of \autoref{def:algoTDA} is in $n^{O_c(1)}$.

Secondly, let $f_t(n,x)$ be the worst case running time of \autoref{def:algoTDB}, where $|V(G)| \le n$, $|X| \le x$, $n \leq |X_0|^{f(c)}$, and $x \leq |X_0|$.
Finally, let $f_{h(t,c)}(n,x)$, with $h(t,c) = (t,c,c_0,|X_0|,p_1,p_2,\epsilon')$, be the worst case running time of \autoref{def:algoTDC},
where $|V(G)| \le n$, $|X| \le x$, for any $v \notin X$, $d_X(v) \le t$, and for any connected component $C \in \C(G,X)^{>p_1(t_i)}$, $\td(C) \le c$.
Observe that $f_t$ and $f_{h(t,c)}$ are non decreasing in each parameter. 
Let $(G,X,k)$ be a worst case running time instance for $B_t(G,X,k)$. 

Notice, re-using notation of the algorithm, that $|N(v)| > t$ . Thus, we get 
\begin{eqnarray*}
f_t(n,x) &= &\max(n^{\O(1)} + f_t(n-1,x) + f_t(n-t,x-t),f_{h(t,c_0)}(n,x))\\
& \le & n^{\O(1)} + f_t(n-1,x) + f_t(n-t,x-t)+f_{h(t,c_0)}(n,x)
\end{eqnarray*}

This leads to $f_t(n,x) \le n^{\O(1)}+nf_{h(t,c_0)}(n,x) + nf_t(n,x-t)$
and then to $f_t(n,x) = {n}^{\O(\frac{x}{t})}f_{h(t,c_0)}(n,x)$ 
This implies that, for any $(n,x)$, we get (assuming $n \leq |X_0|^{f(c)}$ and $x \leq |X_0|$)
\begin{eqnarray*}
    f_t(n,x) & = & {n}^{\O(\frac{x}{t})}f_{h(t,c_0)}(n,x) \\
    & = & {2}^{\O(\frac{x}{t}\log(n))}f_{h(t,c_0)}(n,x) \\
    & = & {2}^{\O(\log(|X_0|)|X_0|^{1-\epsilon})}f_{h(t,c_0)}(n,x).
\end{eqnarray*}

Let us now turn to $f_{h(t,c)}$. Let $(G,X,k)$ be a worst case running time for $C_{h(t,c)}(G,X,k)$. Since this constitutes the main technical part of the analysis, we proceed in several steps.

\paragraph*{Upper bounding $f_{h(t,c)}$ according to the algorithm definition.}

Let us consider different cases according to the fact that $C_{h(t,c)}$ branches on vertices of large degree (line~\ref{line:algoTDC1}), extends the modulator (line~\ref{line:algoTDC2}), or falls into one of the base cases (line~\ref{line:algoTDCNEW} and line~\ref{line:algoTDC3}).
  
\noindent\textbf{Case 1: $C_{h(t,c)}(G,X,k)$ branches on vertices of large degree (line~\ref{line:algoTDC1}).}
  
Notice, re-using notation of the algorithm, that $|N(v)| > t_i$. Thus, in this case we get $f_{h(t,c)}(n,x) \le f_{h(t,c)}^1(n,x)$ where \[f_{h(t,c)}^1(n,x) = n^{\O(1)} + f_{h(t,c)}^1(n-1,x-1) + f_{h(t,c)}^1(n-t_i,x-t_i).\]

\noindent\textbf{Case 2: $C_{h(t,c)}(G,X,k)$ extends the modulator (line~\ref{line:algoTDC2}).}

We begin with the following observation.

\begin{claim}\label{claim:bound_delta_x}
    If we are not in Case 1, then $t_i \ge \Delta_X(G)$.
\end{claim}

\begin{proof}
    Notice that, at recursive call $i$, every vertex outside the current modulator has at most $t+i$ neighbors in it. Indeed, this holds initially by Algorithm $B$, and each extension adds at most one root per connected component of $G \setminus X$. In other words, every vertex remaining outside the modulator gains at most one neighbor in $X$ each time.
    
    On the other hand, if $|X_0|$ is large enough, we have $t+i \le t_i$. Hence, every vertex outside $X$ has at most $t_i$ neighbors in $X$. Finally, since we are not in Case 1, every vertex in $X$ also has at most $t_i$ neighbors in $X$. Therefore, we obtain $t_i \ge \Delta_X(G)$.
\end{proof}

We continue by proving a simple consequence of \autoref{def:boundedLNC} under the assumption that $p_2(y) = \O(y^{\lambda_2})$.

\begin{claim}\label{claim:bound_extention}
    Let $G$ be a graph with $(p_1,p_2)$-bounded large neighborhood components, and let $V(G) = R \cup X$ be a partition with $\td(G[R]) \leq \td$. Suppose that $p_2(y) \leq \alpha_2 y^{\lambda_2}$ for some constant $\alpha_2$. Then, for any $t \geq \Delta_X(G)$ such that $|X| > Lt^5$, it holds that $|\mathcal{C}^{>p_1(t)}(G,X)| \leq \alpha_2t^{\lambda_2}g(\td)|X|$.
\end{claim}

\begin{proof}

Since $t \geq \Delta_X(G)$ and $|X| > Lt^5$, we have

\begin{eqnarray*}
    |\mathcal{C}^{>p_1(t)}(G,X)| & \leq & |\mathcal{C}^{>p_1(\Delta_X(G))}(G,X)| \\
    & \leq & p_2(\Delta_X(G))g(\td)|X| \\
    & \leq & p_2(t)g(\td)|X| \\
    & \leq & \alpha_2t^{\lambda_2}g(\td)|X|.
\end{eqnarray*}

\end{proof}
  
For each $i$, define $z_i(x) = \alpha_2t_i^{\lambda_2}g(c)x$. Notice, re-using notation of the algorithm, that $|Z| \leq |\mathcal{C}^{>p_1(t_i)}(G,X)|$. Since we are not in Case 1, by \autoref{claim:bound_delta_x}, we have $t_i \geq \Delta_X(G)$, and since we did not stop at line~\ref{line:algoTDCNEW}, we have $|X| > Lt_i^5$. Thus, by \autoref{claim:bound_extention}, in this case we get $f_{h(t,c)}(n,x) \le f_{h(t,c)}^2(n,x)$ where \[f_{h(t,c)}^2(n,x) = n^{\O(1)} + f_{h(t+1,c-1)}^2(n,x+z_i(x)).\]

\noindent\textbf{Case 3: $C_{h(t,c)}(G,X,k)$ falls into a base case (lines~\ref{line:algoTDCNEW} and~\ref{line:algoTDC3}).}

First, suppose that $|X| \le Lt_i^5$ and we reach line~\ref{line:algoTDCNEW}. Since $\td(G\setminus X) \le c_0$, we have $\tw(G)\le |X|+c_0=\O(t_i^5)$. Therefore, the instance can be solved in time $n^{\O(1)}2^{\O(t_i^5)}$.

Otherwise, we reach line~\ref{line:algoTDC3}. Since we are not in Case 1, we have $\Delta(G[X]) \le t_i$, and therefore $|E(G[X])| \leq |X|t_i$. On the other hand, suppose that $p_1(y) \leq \alpha_1 y^{\lambda_1}$ for some constant $\alpha_1$. As we are not in Case 2, it follows that $|N_X(C)| \leq p_1(t_i) \leq \alpha_1t_i^{\lambda_1}$ for every connected component $C$ of $G[R]$. Let $\Delta = \alpha_1t_i^{\lambda_1}$. If $\Delta > |X|^{\epsilon_r}$, then as $\Delta = \O(t_i^{\lambda_1})$, we obtain $|X| = \O(t_i^{\lambda_1/\epsilon_r})$, and as $\tw(G) \leq |X| + c_0$, we obtain $\tw(G)= \O(t_i^{\lambda_1/\epsilon_r})$. Otherwise ($\Delta \leq |X|^{\epsilon_r}$), we can apply \autoref{def:degreeTW}, and obtain that
\begin{eqnarray*}
\tw(G) & = & \O(c_0+|X|^{\epsilon_E}t_i^{\epsilon_E} + |X|^{1-\epsilon_{\Delta}}t_i^{\lambda_1\epsilon_{\Delta}}) = \O(|X|^{\epsilon_E}t_i^{\epsilon_E} + |X|^{1-\epsilon_{\Delta}}t_i^{\lambda_1\epsilon_{\Delta}}).
\end{eqnarray*}

Thus, in either base case we get $f_{h(t,c)}(n,x) \le f_{h(t,c)}^3(n)$ where \[f_{h(t,c)}^3(n) = n^{\O(1)}2^{\tw(G)} = n^{\O(1)}2^{\O(t_i^5 + t_i^{\lambda_1/\epsilon_r} + |X|^{\epsilon_E}t_i^{\epsilon_E} + |X|^{1-\epsilon_{\Delta}}t_i^{\lambda_1\epsilon_{\Delta}})}.\] Thus, we obtain $f_{h(t,c)}(n,x) \leq \max(f_{h(t,c)}^1(n,x),f_{h(t,c)}^2(n,x),f_{h(t,c)}^3(n))$.

\paragraph*{Analysis of $f_{h(t,c)}$.}

Let us now bound $f_{h(t,c)}$ by taking the maximum over the three cases, which simplifies to the sum of three values. In fact,
\begin{eqnarray*}
    f_{h(t,c)}(n,x) & \leq & \max(f_{h(t,c)}^1(n,x),f_{h(t,c)}^2(n,x),f_{h(t,c)}^3(n))\\
    & \leq & f_{h(t,c)}^1(n,x) + f_{h(t,c)}^2(n,x) + f_{h(t,c)}^3(n)\\
    & \leq & n^{\O(1)} + f_{h(t,c)}^1(n-1,x-1) + f_{h(t,c)}^1(n-t_i,x-t_i)\\
    & + & f_{h(t+1,c-1)}^2(n,x+z_i(x)) + f_{h(t,c)}^3(n),
  \end{eqnarray*}
implying $f_{h(t,c)}(n,x) = \O(f_{h(t,c)}(n,x-1) + f_{h(t,c)}(n,x-t_i) + f_{h(t+1,c-1)}(n,x+z_i(x)) + f_{h(t,c)}^{3}(n))$. This leads to $f_{h(t,c)}(n,x) = \O(x(f_{h(t,c)}(n,x-t_i) + f_{h(t+1,c-1)}(n,x+z_i(x)) + f_{h(t,c)}^{3}(n)))$, and then to $f_{h(t,c)}(n,x) = \O(x^{x/t_i}(f_{h(t+1,c-1)}(n,x+z_i(x)) + f_{h(t,c)}^{3}(n)))$.

To continue the analysis of $f_{h(t,c)}$, and since Algorithm $C$ is always executed immediately after Algorithm $B$, we assume that we start with the meta parameters produced by Algorithm $A$ and Algorithm $B$. In particular, we have $|X_0|$ = $|X|$ and $c_0 = c$, which also implies we start with $t_0 = |X_0|^{\epsilon'}$. Moreover, the initial value $x$ is upper bounded by $|X_0|$. Now, we define the sequence $x_{i+1} = x_i + z_i$ with $x_0 = x$, where $z_i = z_i(x_i) = \alpha_2t_i^{\lambda_2}g(c)x_i$. With this notation, we can now bound as $f_{h(t_0,c_0)}(n,x) = \O(x_0^{x_0/t_0}(f_{h(t_0+1,c_0-1)}(n,x_0+z_0) + f_{h(t_0,c_0)}^{3}(n)))$, which leads to $f_{h(t_0,c_0)}(n,x) = \O(\prod_{i=0}^{c-1}x_i^{x_i/t_i})f_{h(t_0+c_0,c_0)}^{3}(n)$.

We now control the growth of the sequence $(x_i)_i$. The following claim gives an upper bound in terms of $x_0$ and $t_i$.

\begin{claim}\label{claim:bound_xi}
    For any $i \geq 1$, it holds that $x_i = \O(x_0t_{i-1}^{\lambda_2i})$. 
\end{claim}

\begin{proof}
    Straightforward from the fact that $x_i = x_0\prod_{j=0}^{i-1}(1 + \alpha_2t_{j}^{\lambda_2}g(c))$, which can be proved by induction in $i$. 
\end{proof}

The following claim describes the growth of the sequence $(t_i)_i$.

\begin{claim}\label{claim:expression_ti}
    For any $i \geq 0$, there exists a constant $M_i$ such that $t_i = |X_0|^{M_i\epsilon'}$.
\end{claim}

\begin{proof}
    Straightforward from the fact that $t_i = |X_0|^{(\sum_{j=0}^{i} \frac{i!}{(i - j)!}\lambda_2^j)\epsilon'}$, which can be proved by induction in $i$. 
\end{proof}

Now, observe that the sequence $(t_i)_i$ grows sufficiently fast, which will be crucial for bounding the running time. In particular, using the definition of $t_i$, it follows that for any $i \geq 1$, $t_{i-1}^{\lambda_2i}/t_i = |X_0|^{-\epsilon'}$. This implies that, setting $x = x_0 = |X_0|$ and using \autoref{claim:bound_xi} and \autoref{claim:expression_ti}, we obtain
  \begin{eqnarray*}
    f_{h(t_0,c_0)}(n,|X_0|) & = & \O(\prod_{i=0}^{c-1}x_i^{x_i/t_i})f_{h(t_0+c_0,c_0)}^{3}(n)\\
    & = & \prod_{i=0}^{c-1}x_i^{\O(x_i/t_i)}f_{h(t_0+c_0,c_0)}^{3}(n)\\
    & = & \prod_{i=0}^{c-1}x_i^{\O(|X_0|t_{i-1}^{\lambda_2i}/t_i)}f_{h(t_0+c_0,c_0)}^{3}(n)\\
    & = & \prod_{i=0}^{c-1}x_i^{\O(|X_0|^{1-\epsilon'})}f_{h(t_0+c_0,c_0)}^{3}(n)\\
    & = & \prod_{i=0}^{c-1}2^{\O(|X_0|^{1-\epsilon'}\log(x_i))}f_{h(t_0+c_0,c_0)}^{3}(n)\\
    & = & 2^{\O(|X_0|^{1-\epsilon'}\log(|X_0|))}f_{h(t_0+c_0,c_0)}^{3}(n).
  \end{eqnarray*}
 
To conclude, we derive a bound for $f_{h(t_0+c_0,c_0)}^{3}$.

\begin{claim}\label{claim:bound:tw}
    We have $f_{h(t_0+c_0,c_0)}^{3}(n) = n^{\O(1)}2^{\O(|X_0|^{f(\lambda_1,\lambda_2,\epsilon_r,\epsilon_{\Delta},\epsilon_E,\epsilon')})}$, where \[f(\lambda_1,\lambda_2,\epsilon_r,\epsilon_{\Delta},\epsilon_E,\epsilon') = \max(5M_c\epsilon',M_c\lambda_1\epsilon'/\epsilon_r,\epsilon_E+M_c(1+\lambda_2c)\epsilon_E\epsilon',(1-\epsilon_{\Delta})+M_c(\lambda_1\epsilon_{\Delta}+\lambda_2c(1-\epsilon_{\Delta}))\epsilon').\]
\end{claim}

\begin{proof}

We start from the bound obtained for $\tw(G)$. Using \autoref{claim:bound_xi} and \autoref{claim:expression_ti}, we have
\begin{eqnarray*}
    \tw(G) & = & \O(t_c^5 + t_c^{\lambda_1/\epsilon_r} + (|X_0|t_c^{\lambda_2c})^{\epsilon_E}t_c^{\epsilon_E} + (|X_0|t_c^{\lambda_2c})^{(1-\epsilon_{\Delta})}t_c^{\lambda_1\epsilon_{\Delta}})\\
    & = & \O(t_c^5 + t_c^{\lambda_1/\epsilon_r} + |X_0|^{\epsilon_E}t_c^{(1+\lambda_2c)\epsilon_E} + |X_0|^{(1-\epsilon_{\Delta})}t_c^{\lambda_1\epsilon_{\Delta}+\lambda_2c(1-\epsilon_{\Delta})})\\
    & = & \O(|X_0|^{5M_c\epsilon'} + |X_0|^{M_c\lambda_1\epsilon'/\epsilon_r} + |X_0|^{\epsilon_E+M_c(1+\lambda_2c)\epsilon_E\epsilon'} + |X_0|^{(1-\epsilon_{\Delta})+M_c(\lambda_1\epsilon_{\Delta}+\lambda_2c(1-\epsilon_{\Delta}))\epsilon'}).
\end{eqnarray*}

Therefore, defining $f(\lambda_1,\lambda_2,\epsilon_r,\epsilon_{\Delta},\epsilon_E,\epsilon') = \max(5M_c\epsilon',M_c\lambda_1\epsilon'/\epsilon_r,\epsilon_E+M_c(1+\lambda_2c)\epsilon_E\epsilon',(1-\epsilon_{\Delta})+M_c(\lambda_1\epsilon_{\Delta}+\lambda_2c(1-\epsilon_{\Delta}))\epsilon')$, we deduce $f_{h(t_0+c_0,c_0)}^{3}(n) = n^{\O(1)}2^{\O(|X_0|^{f(\lambda_1,\lambda_2,\epsilon_r,\epsilon_{\Delta},\epsilon_E,\epsilon')})}$.

\end{proof}

Consequently, we obtain \[f_{h(t_0,c_0)}(n,|X_0|) = n^{\O(1)}2^{\O(\log(|X_0|)|X_0|^{\max(f(\lambda_1,\lambda_2,\epsilon_r,\epsilon_{\Delta},\epsilon_E,\epsilon'),1-\epsilon')})}.\]

\paragraph*{Putting the pieces together.}

Finally, we combine the bounds obtained for Algorithms $A$, $B$, and $C$, to derive the overall running time. Given an initial instance $(G_0,X_0,k_0)$, the total running time is
\[
n^{\O(1)}+f_{|X_0|^\epsilon}(n,|X_0|)
  = n^{\O(1)} {2}^{\O(\log(|X_0|)|X_0|^{1-\epsilon})}f_{h(t_0,c_0)}(n,|X_0|) \]
leading to
\[ n^{\O(1)}{2}^{\O(\log(|X_0|)|X_0|^{1-\epsilon})}2^{\O(\log(|X_0|)|X_0|^{\max(f(\lambda_1,\lambda_2,\epsilon_r,\epsilon_{\Delta},\epsilon_E,\epsilon'),1-\epsilon')})}.
\]
Equivalently,
\[
n^{\O(1)}2^{\O(\log(|X_0|)|X_0|^{\max(f(\lambda_1,\lambda_2,\epsilon_r,\epsilon_{\Delta},\epsilon_E,\epsilon'),1-\epsilon,1-\epsilon')})}.
\]

It remains to choose $\epsilon$ and $\epsilon'$ that minimizes $\max(f(\lambda_1,\lambda_2,\epsilon_r,\epsilon_{\Delta},\epsilon_E,\epsilon'),1-\epsilon,1-\epsilon')$. By definition of $f$, it corresponds to minimizing
\[
\max \left\{
5M_c\epsilon',\;
M_c\lambda_1\epsilon'/\epsilon_r,\;
\epsilon_E+M_c(1+\lambda_2c)\epsilon_E\epsilon',\;
(1-\epsilon_{\Delta})+M_c(\lambda_1\epsilon_{\Delta}+\lambda_2c(1-\epsilon_{\Delta}))\epsilon',\;
1-\epsilon,\;
1-\epsilon'
\right\}.
\]
Setting $\epsilon' = \epsilon$, we get that, for $a = M_c(1+\lambda_2c)\epsilon_E$ and $b = M_c(\lambda_1\epsilon_{\Delta}+\lambda_2c(1-\epsilon_{\Delta}))$, we minimize $\max(u_1(\epsilon),u_2(\epsilon),u_3(\epsilon),u_4(\epsilon),u_5(\epsilon))$
where $u_1(\epsilon) = 5M_c\epsilon$, $u_2(\epsilon) = M_c\lambda_1\epsilon/\epsilon_r$, $u_3(\epsilon) = \epsilon_E + a\epsilon$, $u_4(\epsilon) = (1-\epsilon_{\Delta}) + b\epsilon$, $u_5(\epsilon) = 1 - \epsilon$.
Thus, the first four functions are increasing in $\epsilon$, and the last one is decreasing.
For $i \in \{1,2,3,4\}$, let $\epsilon^\star_i$ be such that $u_i(\epsilon^\star_i) = u_5(\epsilon^\star_i)$.
Notice that $\epsilon^\star_1 = \frac{1}{5M_c+1}$, $\epsilon^\star_2 = \frac{\epsilon_r}{M_c\lambda_1+\epsilon_r}$, $\epsilon^\star_3 = \frac{1-\epsilon_E}{a+1} = \frac{1-\epsilon_E}{M_c(1+\lambda_2c)\epsilon_E+1}$, and $\epsilon^\star_4 = \frac{\epsilon_{\Delta}}{b+1} = \frac{\epsilon_{\Delta}}{M_c(\lambda_1\epsilon_{\Delta}+\lambda_2c(1-\epsilon_{\Delta}))+1}$.
Now, observe that $\min \max \{u_1,u_2,u_3,u_4,u_5\}$ is reached for $\epsilon^\star = \min(\epsilon^\star_1,\epsilon^\star_2,\epsilon^\star_3,\epsilon^\star_4)$ 
and has value $1 - \epsilon^\star$.
\end{proof}

Let us now apply the previous theorem on a specific graph class.
Notice that string graphs have $(\frac{1}{4},\frac{1}{2},\frac{1}{6})$-modulator bounded treewidth by \autoref{thm:tw-string-XR}. Together with \autoref{lem:blnc}, it follows that string graphs are a $(p_1,p_2,\frac14,\frac12,\frac16)$ TD-Mod compliant class, where $p_1(x)=\O(x^{2})$ and $p_2(x)=\O(x^{7})$. To make the dependence on $\td$ explicit, we bound $M_{\td}$.

\begin{lemma}\label{lemma:bound_Mc}
  Let $\td \geq 1$, and let $M_{\td}$ be defined as in \autoref{thm:treedepth}, that is, $M_{\td} = \sum_{i=0}^{\td} \frac{(\td)!}{(\td - i)!}\lambda_2^i$. If $\lambda_2 \geq 2$, then $M_{\td} \leq 2 \cdot (\td)! \cdot \lambda_2^{\td}$.
\end{lemma}

\begin{proof}
    Observe that \[M_{\td} = \sum_{i=0}^{\td} \frac{(\td)!}{(\td - i)!}\lambda_2^i = \sum_{i=0}^{\td} \frac{(\td)!}{i!}\lambda_2^{\td-i} = (\td)! \cdot \lambda_2^{\td} \cdot \sum_{i=0}^{\td} \frac{(1/\lambda_2)^{i}}{i!}.\] Since all terms are nonnegative, we have \[\sum_{i=0}^{\td} \frac{(1/\lambda_2)^{i}}{i!} \leq \sum_{i=0}^{\infty} \frac{(1/\lambda_2)^{i}}{i!} = e^{1/\lambda_2}.\] Hence, using $\lambda_2 \geq 2$, we get $e^{1/\lambda_2} \leq e^{1/2} \leq 2$. Therefore, we deduce $M_{\td} \leq 2 \cdot (\td)! \cdot \lambda_2^{\td}$.
\end{proof}

This leads to the following.

\begin{corollary}\label{cor:td}
$\VC/\disto{(td=\O(1))}$ can be solved in time
$n^{\O(1)}2^{|X|^{1-\eta}}$, with $\eta < \frac{1}{2 \cdot (td)! \cdot 7^{\td}(2+35\td)+6}$ in string graphs.
\end{corollary}
\section{Subexponential algorithm for \texorpdfstring{$\VC/\disto{paths}$}{VC/dist-to-paths}}
\label{sec:paths}
\subsection{Algorithm definition}

The algorithm will perform two kinds of branching, one to remove free subpaths (to be defined now), and the other to remove $K_{1,t}$ of large conflict (see \autoref{def:conflict} and \autoref{def:K1T}).
Let us now define free subpaths and provide some properties.

\begin{definition}\label{def:free}
  Given an instance $(G,X,k)$ of VC/dist-to-paths, let $P_0$ be a path in $G \setminus X$.
  Let $P$ be a subpath of $P_0$, with $P=(v_0,\dots,v_\ell)$, where $\ell \ge 0$. We say that \emph{$P$ is a free subpath of $(G,X)$} if there is no $i$ and $j$ such that
  $i$ is even, $j$ is odd, and both $N_X(v_i)$ and $N_X(v_j)$ are non empty. In other words, if $P$ is free, then vertices of $P$ having neighbors in $X$ are either all even or all odd.
  We denote $\delta(P)=\{v_i \in P \mid N_X(v_i) \neq \emptyset\}$, that is, the set of vertices of $P$ that have a neighbor in $X$.
\end{definition}

The purpose of the following definition is to decompose every path in $G \setminus X$ into inclusion wise maximal free subpaths. We construct this decomposition recursively from left to right: starting at the first vertex of a path, we take the longest subpath that is free, we remove it, and we repeat the same procedure on the remaining path. The union of all these inclusion wise maximal free subpaths over all the paths in $G \setminus X$ forms what we call the canonical partition.

\begin{definition}\label{def:canonical}
  Given an instance $(G,X,k)$ of VC/dist-to-paths, we define the \emph{canonical partition $\mathcal{P}^{can}(G,X)$} as follows.
  For any connected component (path) $P$ of $G\setminus X$,
  partition $P$ into a set $\mathcal{P}^{can}(G,X,P)$ of subpaths as follows. Let $P'$ be the inclusion wise maximal free subpath of $P$ starting from the left of $P$. Add
  $P'$ to $\mathcal{P}^{can}(G,X,P)$ and restart with $P \setminus P'$. Now, define the canonical partition $\mathcal{P}^{can}(G,X)$ as the union of $\mathcal{P}^{can}(G,X,P)$ over all connected component $P$ of $G \setminus X$.

  We also define $NC^{can}(G,X)=|\{N_X(P),P \in \mathcal{P}^{can}(G,X), N_X(P) \neq \emptyset\}|$ as the number of distinct non empty neighborhoods in $X$ induced by the subpaths of the canonical partition.
\end{definition}

\begin{lemma}\label{lemma:canonical}
Let $G \in \G$ where $\G$ is a hereditary graph class with $p$-bounded neighborhood complexity.
  Let $(G,X,k)$ be an instance of VC/dist-to-paths.
  \begin{itemize}
  \item For any $t$, if for any $P \in \mathcal{P}^{can}(G,X)$, $d_X(P) \le t$, then for any free subpath $P'$ of $(G,X)$, $d_X(P') \le t$.
  \item $NC^{can}(G,X)\le 2p(|X|,|X|)|X|$.
  \end{itemize}
\end{lemma}
\begin{proof}
  Let $(G,X,k)$ be an instance of VC/dist-to-paths.
  Let $P$ be a connected component (a path) of $G \setminus X$, and let $P'$ be a free subpath of $(G,X)$ contained in $P$. Observe that the vertices $\delta(P')$ must be contained in the vertex set of some path in $\mathcal{P}^{can}(G,X,P)$, implying the first item.

For the second item, notice that we can partition $\mathcal{P}^{can}(G,X)$ into two sets of subpaths $\mathcal{P}^1$ and $\mathcal{P}^2$.
For any $i \in \{1,2\}$, we define $G_i = G[X \cup \bigcup_{P \in \mathcal{P}^i} P]$. As $G_i$ remains in $\G$, $G_i$ has $p$-bounded neighborhood complexity, implying $|\{N_X(P),P \in \mathcal{P}^i\}| \le p(|X|,|X|)|X|$.
\end{proof}

The next lemma shows that a free subpath behaves like a vertex, in the sense that we can always take all even (or odd, depending on which vertices of the free subpath have neighbors in $X$) vertices, or all their neighbors. This will allow us to branch on free subpaths of large degre in $X$. However, branching on one free subpath of large degree at a time
may not lead to a subexponenial running time, just because the number of such free subpaths might not be polynomial in $|X|$. Thus, we will rely on neighborhood complexity
and branch for a whole set of free subpaths having the same large neighborhood, with the idea that this branching decreases $NC^{can}(G,X)$ (which is polynomially
bounded in $|X|$ by \autoref{lemma:canonical}) rather than just $n, k$ or $|X|$.

\begin{lemma}\label{lemma:freeSubpathBranching}
  Let $(G,X,k)$ be an instance of VC/dist-to-paths, $X' \subseteq X$, $X' \neq \emptyset$, $\mathcal{P}(X') = \{P \in \mathcal{P}^{can}(G,X) \mid N_X(P)=X'\}$,
  and $\delta = \bigcup_{P \in \mathcal{P}(X')}\delta(P)$.
  Then,
  \begin{itemize}
  \item $(G,X,k) \Leftrightarrow (G \setminus X', X \setminus X', k - |X'|) \vee (G \setminus \delta, X , k - |\delta|)$,
  \item $NC^{can}(G \setminus \delta, X) \le \max(NC^{can}(G, X)-1,0)$.
  \end{itemize}
\end{lemma}
\begin{proof}
  The only non trivial direction is $\Rightarrow$, so let us assume that $(G,k)$ has a vertex cover $S$ of size at most $k$.
  If $X' \subseteq S$ then $(G \setminus X', X \setminus X', k - |X'|)$ is yes-instance.
  Otherwise, let $v \in X' \setminus S$. For any $P \in \mathcal{P}(X')$, there exists $v_P \in V(P) \cap S$.
  Let $P \in \mathcal{P}(X')$, and let us now prove that we can restructure $S$ into $S'$ such that $|S'| \le |S|$ and that $\delta(P) \subseteq S'$,
  which will imply that $(G \setminus \delta, X , k - |\delta|)$ is a yes-instance.
  
Suppose that there exists $v \in \delta(P) \setminus S$, and let $v$ be such a vertex having a minimum distance from $v_P$.
Let $v_0$ be last vertex in $\delta(P) \cap S$, meaning that in the path from $v_P$ to $v$, we first visit $v_0$, which is the last vertex in $\delta(P) \cap S$,
then we visit vertices $v_1,\dots,v_{2\ell+1},v$, where all $v_i$, $1 \le i \le 2\ell+1$, are not in $\delta(P)$.
As $v \notin S$, this implies that $v_{2\ell+1} \in S$, and that $|S \cap \{v_1,\dots,v_{2\ell+1}\}| \ge \ell+1$.
Thus, we can restructure $S$ by removing vertices of $S \cap \{v_1,\dots,v_{2\ell+1}\}$, and adding instead $v_2,v_4,\dots,v_{2\ell},v$,
and the obtained solution now contains $v$ and is not larger.
Repeating this argument implies that we can build $S'$ as claimed.

Finally, let us prove that $NC^{can}(G \setminus \delta, X) = NC^{can}(G, X)-1$ by proving the following property:
for any path $P \in G \setminus X$, $\{N_X(P'), P' \in \mathcal{P}^{can}(G \setminus \delta,X,P), N_X(P') \neq \emptyset \} = \{N_X(P'), P' \in \mathcal{P}^{can}(G,X,P), N_X(P') \neq \emptyset \} \setminus X'$.
Let $P \in G \setminus X$.
Assume that $\mathcal{P}(X') \neq \emptyset$, and let $P'$ be the first (when computing $\mathcal{P}^{can}(G,X,P)$, meaning the left most) free subpath of $P$ such that $N_X(P')=X'$.
Notice that in $G \setminus \delta$, all remaining vertices in $P' \setminus \delta$ have no neighbor in $X$, and thus for any maximal subpath $\tilde{P}$ of $P'$
between two vertices in $\delta(P')$, $\tilde{P}$ becomes a new element of $NC^{can}(G \setminus \delta, X)$ with no neighbor in $X$.
Therefore, the result is immediate by induction.

\end{proof}

We are now ready to describe the main algorithm.
Notice that branching to 
obtain $\Delta(G[X]) \le t$ (for a well chosen $t$)
is done as a preprocessing in \autoref{def:algoPathMain}, as
$\Delta(G[X])$ will never increase during \autoref{def:algoPath}.

\begin{algorithm}
\caption{$A_{\epsilon_1,\epsilon_2}(G,X,k)$}
\begin{algorithmic}[1]
  \State let $t=|X|^{\epsilon_1}$ and $c = |X|^{\epsilon_2}$
  \State Apply algorithm of \autoref{lemma:prelimBranching} with $t$ 
  to obtain a set $\I_{\epsilon_1}$ of instances.
  \State return $\bigvee_{(G',X',k') \in \I_{\epsilon_1}} B_{t,c}(G',X',k')$
\end{algorithmic}
\label{def:algoPathMain}
\end{algorithm}

\noindent Recall in the definition of \autoref{def:algoPath} that $\mathcal{P}(X') = \{P \in \mathcal{P}^{can}(G,X) \mid N_X(P)=X'\}$.

\begin{algorithm}
\caption{$B_{t,c}(G,X,k)$}
\begin{algorithmic}[1]
  \Require $\Delta(G[X]) \le t$
 
  \State if $b(G,X,k)<0$ return false, and if $b(G,X,k) \ge |X|$ return true
  \State if there exists $X' \subseteq X$ with $|X'| \ge t$ such that $\mathcal{P}(X') \neq \emptyset$,
       define $\delta = \bigcup_{P \in \mathcal{P}(X')}\delta(P)$, and return $B_{t,c}(G \setminus X', X \setminus X', k - |X'|) \vee B_{t,c}(G \setminus \delta, X , k - |\delta|)$ \label{line:algoPath1}
  \If{there exists $(v,X')$, a $K_{1,c}$ of conflict $c$ (recall \cref{def:K1T})}
    \State return $B_{t,c}(G-v,X-v,k-1) \vee B_{t,c}(G-X',X-X',k-c) \vee $ \\ $\bigvee_{v' \in X'}B_{t,c}(G-N(v)-N(v'),X \setminus (N(v) \cup N(v')), k-|N(v) \cup N(v')|)$ \label{line:algoPath2}
    \Else
    \State Decide $(G,k)$ using a standard DP parameterized by $\tw$ (as $\tw(G)$ is small). \label{line:algoPath3}
  \EndIf
\end{algorithmic}
  \label{def:algoPath}
\end{algorithm}

Correctness of this algorithm is immediate according to \autoref{lemma:freeSubpathBranching} for the branching on free subpaths, and by observing that the branching on $(v,X')$, a $K_{1,c}$ of conflict $c$ is indeed safe: either there exists a $v' \in X'$ such that the solution does not intersect $\{v,v'\}$ (corresponding to the third branch), or the solution has to contain $v$ or $X'$ (corresponding to the first two branches). To bound its running time and decide how to set the parameters $\epsilon_1,\epsilon_2$, we need the lemmas of next sections.

\subsection{Bounding the treewidth when cut outputs a small number of paths}

\begin{figure}[htbp]
    \centering
    \includegraphics[width=0.85\linewidth]{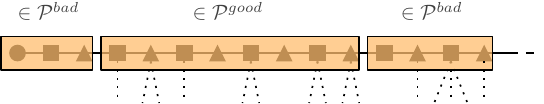}
    \caption{An example of $cut_\Delta(G,X,P)$ where $P$ is an odd path. Shifted-even vertices are represented by squares, and shifted-one ones by triangles. Doted lines represent edges between vertices of $P$ and $X$. The three first path constructed by $cut$ are represented by the three orange rectangles.}
    \label{fig:cut}
\end{figure}

Let us first define the set of sub-paths $cut_\Delta(G,X,P)$ obtained by \autoref{def:cut} (see also \autoref{fig:cut}) that cut paths of large degree in $X$ into subpaths whose degree in $X$ is controlled (but length is arbitrary). 
Moreover, we define $cut_\Delta(G,X)=\bigcup_{P \in \mathcal{C}^{\ge \Delta}} cut_\Delta(G,X,P)$.
 
\begin{algorithm}[!h]
\caption{$cut_\Delta(G,X,P)$}
\label{def:cut}
\begin{algorithmic}[1]
    \Require an instance $(G,X,k)$ of VC/dist-to-paths, an integer $\Delta$, and a connected component (a path)
  $P$ of $G\setminus X$ such that $|N_X(P)| \ge \Delta$
  \State Let $\mathcal{P}^{good} = \mathcal{P}^{bad} = \emptyset$.
  \State Let $P=(v_0,\ldots,v_{\ell})$.
  \State If $|V(P)|$ is even then define $s=0$, otherwise define $s=1$. Let $\{v_{s+2x} \ |\ x \ge 0\}$ be the set of shifted-even vertices of $P$, and $\{v_{s+2x+1}\ |\ x \ge 0\}$ the one of shifted-odd vertices of $P$ (if $s=1$ then $v_0$ is neither a shifted-even nor shifted-odd vertex of $P$).
  \State Let $v_i$ be the minimum shifted-even vertex such that $N_X(v_i) \neq \emptyset$. Add $(v_s,\dots,v_{i-1})$ (which might be empty) to $\mathcal{P}^{bad}$ and set $s=i$. \label{cut:line1}
  \While{it is possible}
    
    \State Let $i>s$ be the minimum shifted-odd vertex $v_i$ such that $P'=(v_s,\dots,v_i)$ has $d_X(P') \ge \Delta$ and $N_X(v_i) \neq \emptyset$. If there is no such $i$ then add $(v_s,\dots,v_{|V(P)|})$ to $\mathcal{P}^{bad}$. Otherwise, add $(v_s,\dots,v_i)$ to $\mathcal{P}^{good}$. 
    \State Define $j>i$ as the minimum shifted-even vertex such that $N_X(v_j) \neq \emptyset$. If there is no such $j$ then add $(v_{i+1},\dots,v_{|V(P)|})$ to $\mathcal{P}^{bad}$ and stop. Otherwise, add $(v_{i+1},\dots,v_{j-1})$ (which might be empty) to $\mathcal{P}^{bad}$, and set $s=j$. \label{cut:line2}
  \EndWhile
  \State output $\mathcal{P}^{good} \cup \mathcal{P}^{bad}$. 

\end{algorithmic}
\end{algorithm}

\begin{lemma}\label{lemma:cut}
Let $(G,X,k)$ be an instance of VC/dist-to-paths, $\Delta$ be an integer, $P$ be a connected component (a path) of $G \setminus X$, and let $\mathcal{P}=cut_\Delta(G,X,P)$.
\begin{enumerate}
\item\label{item:cut1} For any $P' \in \mathcal{P}$, $|V(P')|$ is even, the first vertex of $P'$ is shifted-even, the last one is shifted-odd.
\item\label{item:cut2} For any $P' \in \mathcal{P}^{good}$, the first vertex and last vertices of $P'$ have non empty neighborhood in $X$.
\item\label{item:cut-nobadbad} If $P$ is cut in more than two paths, there are never two consecutive bad ones.
\item\label{item:cut3} $\vc(P)=\sum_{P' \in \mathcal{P}}\vc(P')$.
\item\label{item:cut4} For any $t$, if for any free subpath $P$ in $(G,X)$ we have $d_X(P) \le t$, then:
  \begin{itemize}
    \item for any $P' \in \mathcal{P}^{good}$, $\Delta \le d_X(P') \le \Delta+2t$,
    \item for any $P' \in \mathcal{P}^{bad}$, $d_X(P') \le \Delta+2t$.
  \end{itemize}
\end{enumerate}
\end{lemma}
\begin{proof}
  The first three items directly follow from the definition of $cut$. The \autoref{item:cut3} follows from the fact that all $|V(P')|$ are even.

  For \autoref{item:cut4}, let us start with a $P' \in \mathcal{P}^{good}$, and let $P'=(v_s,\dots,v_i)$.
  If there is no shifted-odd vertex $i'$ with $s < i' < i$ such that $N_X(v_{i'}) \neq \emptyset$, then $(v_s,\dots,v_{i-1})$ is a free-subpath, implying that
   $d_X(P') \le d_X(v_s,\dots,v_{i-1})+d_X(v_i) \le 2t$.
  Otherwise, let $i'$ the largest shifted-odd vertex $i'$ with $s < i' < i$ such that $N_X(v_{i'}) \neq \emptyset$. By definition of $i$, $d_X(v_s,\dots,v_{i'}) \le \Delta$.
  Moreover, by definition of $i'$, all shifted-odd vertices between $i'$ and $i$ have no neighbor in $X$, implying that $(v_{i'+1},\dots,v_{i-1})$ is a free-subpath, implying that
  $d_X(P') \le d_X(v_s,\dots,v_{i'})+d_X(v_{i'+1},\dots,v_{i-1})+d_X(v_i) \le \Delta+2t$.
  
  Let us now consider a $P' \in \mathcal{P}^{bad}$. If $P'$ was added in line~\ref{cut:line1} or~\ref{cut:line2}, then $P'$ is a free subpath, implying $d_X(P') \le t$.
  Otherwise, let  $P'=(v_s,\dots,v_{|V(P)|})$ for some shifted-even $s$ such that there is no $i>s$ shifted-odd vertex $v_i$ such that $d_X(v_s,\dots,v_i) \ge \Delta$ and $N_X(v_i) \neq \emptyset$. 
  If there is no $i>s$ such that $d_X(v_s,\dots,v_i) \ge \Delta$ then we are done, and otherwise let $i>s$ be minimum such that $d_X(v_s,\dots,v_i) \ge \Delta$, implying that $N_X(v_{i'}) \neq \emptyset$,
  and thus that $i'$ is a shifted-even vertex.
  By definition, all shifted-odd vertices $v_{i'}$ for $i' > i$ have $N_X(v_{i'})=\emptyset$, implying that $(v_{i},\dots,v_{|V(P)|})$ is a free subpath.
  Thus, $d_X(P') \le d_X(v_{s},\dots,v_{i-1}) + d_X(v_{i},\dots,v_{|V(P)|}) \le \Delta + t$.
\end{proof}

The next lemma tells us that, if $|\mathcal{P}|$ is almost linear in $X$, then we are done as $\tw(G)=o(|X|)$. Informally, we can chose the parameters $\epsilon, \epsilon_{\Delta}, \epsilon_{cut}$ of the next lemma to be as small as we want,
and typically smaller than the ``fixed'' constant $\epsilon_{\Delta}$ that depends only of the graph class.

\begin{lemma}\label{lemma:tw}
  Let $t$ and $\Delta$ be positive integers.
  Let $\G$ be an $(\epsilon_r,\epsilon_E,\epsilon_\Delta)$-modulator bounded treewidth class, and let $(G,X,k)$ be an instance of VC/dist-to-paths such that $G \in \G$, $\Delta(G[X]) \le t$, and $d_X(P) \le t$ for every free subpath $P$ in $(G,X)$. Let $\mathcal{P}=cut_\Delta(G,X)$ and define $\Delta'= 2\Delta+4t+2$. Assume that $\Delta' \le |X|^{\epsilon_r}$.
        
  Then, $\tw(G)=\O((t(|X|+|\mathcal{P}^{good}|))^{\epsilon_E}+(|X|+|\mathcal{P}^{good}|)^{(1-\epsilon_\Delta)}\Delta'^{\epsilon_\Delta})$.
\end{lemma}

\begin{proof}
  Let $(G,X,k)$ be such an instance. Let $Z$ be the set containing the first vertex of every good path in $\mathcal{P}$,
  and let $X'=X \cup Z$.
  We first observe that, for any connected component $P'$ of $G \setminus X'$, we have $d_{X'}(P') \le \Delta'$.
  Indeed, suppose first that $P'$ is a path of $G \setminus X$ satisfying $d_X(P') < \Delta$. In this case, this path was not given to $cut$ as input. Moreover, since all vertices in $Z$ belong to paths other than $P'$, we have $d_{X'}(P') = d_X(P') \le \Delta'$.
  The same occurs for a path of $G \setminus X$ leading to no good paths. Indeed, in such case it would be the union of at most two bad paths of degree less than $\Delta$.
  
  Suppose next that $P' = \{v_0\}$, where $v_0$ is the first vertex skipped by $cut$ when the original path has odd order. Then, $v_0$ is a free subpath and has at most one neighbor in $Z$. Hence, $d_{X'}(P') \le t+1 \le \Delta'$.
  
  In all remaining cases, there exists $P \in \mathcal{P}^{good}$ such that $P'$ is obtained from $P$, and possibly a bad path following it, and by removing first vertex of $P$. By \autoref{lemma:cut} \autoref{item:cut4}, we get $d_X(P) \le \Delta+2t$, such as for the bad path, and thus $d_X(P') \le 2\Delta+4t$. Moreover, $P'$ has at most two neighbors in $Z$: the first vertex of $P$, and possibly the first vertex of the path in $\mathcal{P}^{good}$ immediately following $P$. Hence, $d_{X'}(P') \le d_{X}(P)+2 \le \Delta'$.

  Next, notice that $|E(G[X'])| \le |E(G[X])|+ |\mathcal{P}^{good}|t$ since every vertex $v \in Z$ is a free subpath,
  leading to $|E(G[X'])| = \O(t(|X|+|\mathcal{P}^{good}|))$.
  
  Thus, as $\Delta' \le |X|^{\epsilon_r} \le |X'|^{\epsilon_r}$, since $\G$ has $(\epsilon_r,\epsilon_E,\epsilon_\Delta)$-modulator bounded treewidth, the definition applied for $X'$ and $\Delta'$ implies the claimed bound on $\tw(G)$.
\end{proof}

\subsection{Finding a \texorpdfstring{$K_{1,c}$}{K1,c} of large conflict when cut outputs a large number of paths}

We will need to prove that, when $cut$ outputs a large number of paths, there exists a $K_{1,c}$ of large conflict, and that during
the branching on this $K_{1,c}$, in the branch where we do not cover a non edge of large conflict, then the budget decreased by a large amount.
Let us start with the following lemma that relates conflict and budget.

\begin{lemma}\label{lemma:conflict_budget}
  Let $(G,X,k)$ be an instance of VC/dist-to-paths and $R = V(G) \setminus X$. Let $X' \subseteq X$ and $c = \conf_R(X')$.
  Let $S$ be a vertex cover of size at most $k$ such that $X' \cap S = \emptyset$. Then, $b(G \setminus N(X'), X \setminus N(X'), k - |N(X')|) \le b(G,X,k)-c$.
\end{lemma}
\begin{proof}
  Let $\{P'_i\}$ be the set of paths obtained from the paths in $R$ after removing the vertices in $N_R(X')$, and let $S^{'*}_i$ be a minimum vertex cover of $P'_i$.
  Since $\conf_R(X') = c$ and $N_R(X') \cup \bigcup_i S^{'*}_i$ is a vertex cover of $G[R]$, we have $|N_R(X')|+\sum_i \vc(P'_i) \ge \vc(G[R])+c$.
  Therefore,
  \begin{eqnarray*}
    b(G \setminus N(X'), X \setminus N(X'), k - |N(X')|) & =& (k-|N(X')|)-\sum_i \vc(P'_i) \\
    & \le & (k-|N(X')|)-(\vc(G[R])+c-|N_R(X')|) \\
    & \le & (k-\vc(G[R]))-c \\
    & = & b(G,X,k)-c.
  \end{eqnarray*}
\end{proof}

Let us now relate paths created by $cut$ and conflict in $X$.

\begin{definition}
 Given an instance $(G,X,k)$ of VC/dist-to-paths and a connected subgraph $P'$ of $G\setminus X$ (a path), we define $\tilde{E}_X(P')$
 as the set of pairs $\{u,v\}\subseteq X$ of non-adjacent vertices such that $\conf_{P'}(B) > 0$.
\end{definition}

We next show that non-free paths of large degree in $X$ generate a lot of pairs in $X$ having a non-zero conflict.
Observe that the technical conditions in the next lemma, namely that $|V(P')|$ is even and that its first and last vertices have neighbors in $X$, are satisfied by paths in $\mathcal{P}^{good}$ created by $cut$ according to \autoref{lemma:cut} \autoref{item:cut1} and \autoref{lemma:cut} \autoref{item:cut2}.

\begin{lemma}\label{lemma:ex}
  Let $(G,X,k)$ be an instance of VC/dist-to-paths, and let $t$ be such that $\Delta(G[X]) \le t$. Let  $P'=(v_0,\dots,v_i)$ be a connected subgraph of $G\setminus X$ (a path) with $|V(P')|$ even, and such that $d_X(v_0)>0$
  and $d_X(v_i)>0$. Then, 
  $\tilde{E}_X(P')$ contains a star with at least $\frac12 d_X(P')-t-1$ leaves.
\end{lemma}
\begin{proof}
We first extend the notion of conflict as follows.
Given $Y \subseteq P'$, let $\conf_{P'}(Y)=|Y|+\vc(P' \setminus Y)-\vc(P')$, which corresponds to the additional cost incurred when the vertices in $Y$ are forced to belong to a vertex cover of $P'$.
  Let $Z = \{\{v_j,v_k\} \subseteq V(P')\ |\ \conf_{P'}(\{v_j,v_k\}) > 0 \}$. Notice that, for any $\ell$, $\{v_0,v_{2\ell+1}\} \in Z$ and $\{v_{2\ell},v_{i}\} \in Z$.
Let $N_{odd} = \bigcup_\ell N_X(v_{2\ell+1})$ and $N_{even} = \bigcup_\ell N_X(v_{2\ell})$.

Fix $u \in N_X(v_0)$. The property of $Z$ above implies that, for any $u' \in N_{odd}$ such that $\{u,u'\} \notin E(G)$ and $u' \neq u$, we have $\{u,u'\} \in \tilde{E}_X(P')$. Indeed, avoiding both $u$ and $u'$ in a vertex cover, forces the inclusion of $v_0$ and some $v_{2\ell+1}$ in $P'$, generating a conflict.
Since $u$ has at most $t$ neighbors in $X$, it follows that $\tilde{E}_X(P')$ contains a star centered at $u$ with at least $|N_{odd}|-t-1$ leaves.
Similarly, fix $v \in N_X(v_i)$. The same argument shows that $\tilde{E}_X(P')$ contains a star centered at $v$ with at least $|N_{even}|-t-1$ leaves.
Since $N_X(P') = N_{odd} \cup N_{even}$, one of these two stars has at least $\frac12 d_X(P')-t-1$ leaves as claimed.
\end{proof}

Let us now define an auxiliary edge-weighted graph $H$ that summarizes the conflict generated by a set of disjoint subpaths.
\begin{definition}\label{def:H}
  Given an instance $(G,X,k)$ of VC/dist-to-paths and a set $\mathcal{P}$ of disjoint connected subgraphs $P'$ of $G\setminus X$,
  we define $H(G,X,\mathcal{P})$ as an edge-weighted undirected graph, with vertex set $X$, such that every $\{u,v\} \subseteq X$ with $\{u,v\} \notin E(G)$ is an edge
  of $H$ with weight $w(\{u,v\}) = |\{P' \in \mathcal{P} \mid \{u,v\} \in \tilde{E}_X(P')\}|$.
  Moreover, given $\Delta > 0$, we define $H_\Delta(G,X)= H(G,X,\mathcal{P}^{good})$, where $\mathcal{P}^{good}$ is computed by $cut_\Delta(G,X)$. 
\end{definition}

\begin{lemma}\label{lemma:conf_weight}
  Let $(G,X,k)$ be an instance of VC/dist-to-paths, let $\Delta$ be an integer, and let $R = V(G) \setminus X$.
Let $w$ be the weight function of $H_\Delta(G,X)$.
Then, for any $\{u,v\} \subseteq X$ such that $\{u,v\} \notin E(G)$, $\conf_R(\{u,v\}) \ge w(\{u,v\})$. 
\end{lemma}
\begin{proof}
  Let $\{u,v\} \subseteq X$ with $\{u,v\} \notin E(G)$. Consider a connected component (a path) $P \in \mathcal{C}^{\ge \Delta}(G,X)$ and let $\mathcal{P}=cut_\Delta(G,X,P)$. By \autoref{lemma:cut} \autoref{item:cut3}, \[\vc(P)=\sum_{P'\in\mathcal{P}}\vc(P').\] 
  Observe that the paths in $\mathcal{P}$ are vertex-disjoint and partition $V(P)$, except possibly for the first vertex when $P$ is odd. Hence, \[|N_P(\{u,v\})| \geq \sum_{P'\in\mathcal{P}}|N_{P'}(\{u,v\})|.\]
  The graph $P\setminus N_P(\{u,v\})$ containing as a subgraph the disjoint union of the graphs $P'\setminus N_{P'}(\{u,v\})$, for $P'\in\mathcal{P}$,
  \[\vc(P\setminus N_P(\{u,v\})) \ge \sum_{P'\in\mathcal{P}}\vc(P'\setminus N_{P'}(\{u,v\})).\] It follows that, \[\conf_P(\{u,v\}) \ge \sum_{P'\in\mathcal{P}}\conf_{P'}(\{u,v\}).\]
  Summing over all components in $\mathcal{C}^{\ge\Delta}(G,X)$ gives \[\conf_R(\{u,v\}) \ge \sum_{P\in cut_\Delta(G,X)}\conf_P(\{u,v\}).\] By definition of $w$, there are exactly $w(\{u,v\})$ paths $P\in\mathcal{P}^{good}$ such that $\conf_P(\{u,v\})>0$. Therefore, \[\conf_R(\{u,v\}) \ge \sum_{\substack{P\in\mathcal{P}^{good}\\ \conf_P(\{u,v\})>0}} \conf_P(\{u,v\}) \ge w(\{u,v\}),\] implying the result.
\end{proof}

\begin{corollary}\label{cor:conf_weight}
  Let $(G,X,k)$ be an instance of VC/dist-to-paths, let $\Delta$ be an integer, and let $R = V(G) \setminus X$. Let $w$ be the weight function of $H_\Delta(G,X)$ and let $c$ be an integer. If $H_\Delta(G,X)$ contains a $K_{1,c}$, denoted by $K=(v,\{v_1,\dots,v_c\})$, such that $w(\{v,v_i\}) \ge c$ for any $i \in [c]$, then $K$ is a $K_{1,c}$ of conflict $c$ in $(G,X)$.
\end{corollary}

\begin{lemma}\label{lemma:cutlarge}
  Let $t$ and $\Delta$ be integers such that $\Delta \ge 4t+4$.
  Let $\G$ be a hereditary graph class with $p$-bounded neighborhood complexity. 
  Let $(G,X,k)$ be an instance of VC/dist-to-paths such that $G \in \G$, $\Delta(G[X]) \le t$, and $d_X(P) \le t$ for any free subpath $P$ of $(G,X)$.
  
  Then, for any integer $w$, if the set $\mathcal{P}^{good}$ computed by $cut_\Delta(G,X)$ has size 
  $|\mathcal{P}^{good}| \ge \frac92 w \cdot \Delta^2 \cdot p(t,\frac32 \Delta)|X|$, then there exists a $K_{1,\frac14 \Delta}$ of conflict $w$ in $(G,X)$.
\end{lemma}
\begin{proof}
  Let $\mathcal{P}'$ be an inclusion-wise maximal subset of $\mathcal{P}^{good}$ such that there is no edge between any two subpaths of $\mathcal{P}'$. 
  Observe that $|\mathcal{P}'| \ge \frac{1}{2}|\mathcal{P}^{good}|$ since, whenever $\mathcal{P}^{good}$ contains several consecutive subpaths of the same path, we can select one out of every two such subpaths. 
  
  By \autoref{lemma:cut} \autoref{item:cut4}, for any $P \in \mathcal{P}'$, $d_X(P) \le \Delta+2t \le \frac32 \Delta$. 
  Let $G'=G[X \cup \bigcup_{P \in \mathcal{P}'}P]$.
  Since $G' \in \G$, $G'$ has $p$-bounded neighborhood complexity, 
  implying $|\{N_X(P), P \in \mathcal{P}'\}| \le p(t,\frac32 \Delta)|X|$.
  This implies that, for any $x$, if $|\mathcal{P}^{good}| \ge 2xp(t,\frac32 \Delta)|X|$, then there exists $X' \subseteq X$ and a subset $\mathcal{P}^* \subseteq \mathcal{P}'$ of size $x$, such that for any $P \in \mathcal{P}^*$, $N_X(P)=X'$. 
  We now show how this implies the existence of a large star of large conflict. We first need the following claim.

    \begin{claim}\label{claim:ballsandbin}
      Let $[D]$ be a set of $D$ colors. Let $0 < \ell \le D$, and let $E_i \subseteq [D]$, $1 \le i \le y$, be subsets of colors such that $|E_i|=\ell$ for any $i$. Let $w > 0$ be an integer.
      If $y > D(w-1)$, then there exists a subset of colors $C \subseteq [D]$ such that $|C|=\ell$ and, for any $c \in C$, $|\{i \mid c \in E_i\}| \ge w$. 
    \end{claim}
    \begin{proof}
      Suppose that no such subset of colors $C$ exists, and let us prove that $y \le D(w-1)$.
      Let $C^+ = \{c \in [D] : |\{i \mid c \in E_i\}| \ge w\}$ and let $C^- = [D] \setminus C^+$.
      We know that $|C^+| \le \ell-1$. Since $|E_i|=\ell$ for any $i$,
      it follows that, for any $i$, there exists a color $c_i \in E_i\cap C^-$. Moreover, every color in $C^-$ appears in at most $w-1$ sets $E_i$. Since $|C^-|\le D$, we obtain
      $y \le D(w-1)$, implying the result.
    \end{proof}
    As $\Delta(G[X]) \le t$, \autoref{lemma:ex} implies that, for any $P \in \mathcal{P}^*$, $\tilde{E}_X(P)$ contains a star $K(P)=(v_{P},X_{P})$ with $|X_{P}| \ge \frac12 |X'|-t-1$ leaves. 
    Since $|X'|\ge \Delta \ge 4t+4$, $|X_{P}| \ge \frac12 |X'| - \frac14 |X'| \ge \frac14 |X'|$. Thus, for each $P \in \mathcal{P}^*$, we may assume that $|X_P|=\frac14|X'|$ by keeping an arbitrary subset of $X_P$ of this size.
Let us now consider a $v \in X'$ that maximizes the number of paths $P \in \mathcal{P}^*$ such that $v_P=v$. Let $\mathcal{P}^*_v = \{P \in \mathcal{P}^* \mid v_P = v\}$ and $y=|\mathcal{P}^*_v|$.
Notice that, by the pigeonhole principle, $y \ge \frac{x}{|X'|}$. We can now apply \autoref{claim:ballsandbin} with the colors in $X'\setminus v$ and the sets $E_i$ corresponding to the sets $X_P$ for $P \in \mathcal{P}^*_v$.
Thus, if $y > D(w-1)$, where $D=|X'|-1$, then the set of colors $C$ given by the claim forms, together with $v$, a star $(v,C)$ of weight at least $w$.
In particular, if $x \ge |X'|^2w$, there exists a star $K_{1,\frac14 |X'|}$ of conflict $w$ in $(G,X)$.
Since $\Delta \le |X'| \le \frac32 \Delta$, if $x \ge \frac94 \Delta^2 w$, then there exists a star $K_{1,\frac14 \Delta}$ of conflict $w$ in $(G,X)$.
\end{proof}

\subsection{Running time analysis}

\begin{theorem}\label{thm:path}
  Let $\G$ be a PATH-Mod compliant class with parameters $(p,\epsilon_r,\epsilon_E,\epsilon_\Delta)$ where $p(x,y)=\O((xy)^\lambda)$.
  Then, $\VC/\disto{\text{paths}}$ restricted to graphs in $\G$ can be solved in parameterized subexponential time.
  More precisely, we can solve an instance $(G,X,k)$ of $\VC/\disto{\text{paths}}$ in time $n^{\O(1)}2^{\log(|X|)|X|^{1-\epsilon^\star}}$, where \[\epsilon^\star = \min(\frac{\epsilon_r}{\epsilon_r+1},\frac{1-\epsilon_E}{(2\lambda+5)\epsilon_E+1},\frac{\epsilon_{\Delta}}{2(\lambda+2)(1-\epsilon_{\Delta})+\epsilon_{\Delta}+1}).\]
\end{theorem}
\begin{proof}

  Let  $(G_0,X_0,k_0)$ be an instance as considered in the theorem with $n_0=|V(G_0)|$. Let $\epsilon_2 \ge \epsilon_1 > 0$ that we will fix later.
  Following notation of \autoref{def:algoPathMain}, let $t=|X_0|^{\epsilon_1}$, and $c=|X_0|^{\epsilon_2}$.
  According to \autoref{lemma:prelimBranching}, we obtain in time $n_0^{\O(1)}2^{\O(\log(|X_0|)|X_0|^{1-\epsilon_1})}$ a set of instances
  $\I_{\epsilon_1}$ with $|\I_{\epsilon_1}|=2^{\O(\log(|X_0|)|X_0|^{1-\epsilon_1})}$.
  Notice that in any instance $(G,X,k) \in \I_{\epsilon_1}$, $G$ is an induced subgraph of $G_0$, and for any $v \in X$, $d_{X}(v) \le t$.

  Let $g_{t,c}(n,nc,b)$ be the worst case running time of \autoref{def:algoPath}, where $V(G) \le n$, $NC^{can}(G,X) \le nc$, $b(G,X,k) \le b$, and for any $v \in X$, $d_{X}(v) \le t$. Observe that $g$ is non decreasing in each parameter.
  Let $(G,X,k)$ be a worst case running time for $B_{t,c}(G,X,k)$.

\paragraph*{Upper bounding $g_{t,c}$ according to the algorithm definition.}
  \begin{claim}\label{claim:bound}
    For any instance $(G,X,k)$, and any recursive call $(G',X',k')$ of $B_{t,c}(G,X,k)$, \autoref{lemma:canonical} implies that $NC^{can}(G',X') \le 2p(|X'|,|X'|)|X'| \le 2p(|X_0|,|X_0|)|X_0|$.
  \end{claim}

  Let us consider different cases according to the fact that $B_{t,c}$ branches on free subpaths (line~\ref{line:algoPath1}), branches on a $K_{1,t'}$ (line~\ref{line:algoPath2}), or fall into the base case (line~\ref{line:algoPath3}).

\noindent\textbf{Case 1: $B_{t,c}(G,X,k)$ branches on free subpaths (line~\ref{line:algoPath1}).}
  
  Notice, re-using notation of the algorithm, that $b(G \setminus X', X \setminus X', k - |X'|) \le b(G,X,k)-t$  and by \autoref{lemma:freeSubpathBranching}, $NC^{can}(G \setminus \delta, X) \le NC^{can}(G,X) -1$ as in that case $NC^{can}(G,X) \geq 1$.
  Thus, by \autoref{claim:bound}, in this case we get
  $g_{t,c}(n,nc,b) \le g_1(n,nc,b)$  where \[g_1(n,nc,b) =n^{\O(1)}+g_{t,c}(n,2p(|X_0|,|X_0|)|X_0|,b-t)+g_{t,c}(n,nc-1,b).\]

  \noindent\textbf{Case 2: $B_{t,c}(G,X,k)$ branches on $K_{1,t'}$ (line~\ref{line:algoPath2}).}
  
  Notice, re-using notation of the algorithm, that $b(G-v,X-v,k-1) \le b(G,X,k)-1$, $b(G-X',X-X',k-c) \le b(G,X,k)-c$, and for any $v' \in X'$,
  by \autoref{cor:conf_weight} and \autoref{lemma:conflict_budget}, $b(G-N(v)-N(v'),X \setminus (N(v) \cup N(v')), k-|N(v) \cup N(v')|) \le b(G,X,k)-c$. Thus, by \autoref{claim:bound}, in this case we get
  $g_{t,c}(n,nc,b) \le g_2(n,nc,b)$ where 
  \[g_2(n,nc,b) = n^{\O(1)}+g_{t,c}(n,2p(|X_0|,|X_0|)|X_0|,b-1)+(c+1)g_{t,c}(n,2p(|X_0|,|X_0|)|X_0|,b-c). \]

  \noindent\textbf{Case 3: $B_{t,c}(G,X,k)$ falls into the base case (line~\ref{line:algoPath3}).}
  
  As we are not in Case 1, we know that for any $P' \in \mathcal{P}^{can}(G,X)$, $d_X(P')\le t$. By \autoref{lemma:canonical}, it implies that for any free subpath $P''$ of $(G,X)$,
  $d_X(P'') \le t$.  
  By \autoref{lemma:cutlarge} with 
  $\Delta=4c$ and $w=c$ (notice that, as required, $\Delta \ge 4t+4$ 
  because $\epsilon_2 \ge \epsilon_1$), as there is no $K_{1,c}$ of conflict $c$ in $(G,X,k)$, then $|{\mathcal P}^{good}| =  \O(c^3(tc)^\lambda|X|) =\O(|X_0|^{1+\epsilon_{cut}})$, 
  for $\epsilon_{cut} \le \lambda\epsilon_1+(\lambda+3)\epsilon_2$.
  Let $\Delta'=2\Delta+4t+2$. If $\Delta' > |X|^{\epsilon_r}$, then as $\Delta' = \O(|X_0|^{\epsilon_2})$, we obtain $|X| = \O(|X_0|^{\frac{\epsilon_2}{\epsilon_r}})$,
  and as $\tw(G) \le |X|+1$, we obtain $\tw(G)= \O(|X_0|^{\frac{\epsilon_2}{\epsilon_r}})$.
  Otherwise ($\Delta' \le |X|^{\epsilon_r}$), we can apply \autoref{lemma:tw}, and obtain that
  \begin{eqnarray*}
    \tw(G) &= & \O(|X_0|^{(1+\epsilon_{cut}+\epsilon_1)\epsilon_E}+|X_0|^{(1+\epsilon_{cut})(1-\epsilon_\Delta)+(\epsilon_2\epsilon_\Delta)}).
  \end{eqnarray*}
Considering both cases, we have $\tw(G) = \O(|X_0|^{f(\lambda,\epsilon_\Delta,\epsilon_E,\epsilon_1,\epsilon_2)})$
where $f(\lambda,\epsilon_r,\epsilon_\Delta,\epsilon_E,\epsilon_1,\epsilon_2)= \max(\frac{\epsilon_2}{\epsilon_r},(1+\epsilon_{cut}+\epsilon_1)\epsilon_E,(1+\epsilon_{cut})(1-\epsilon_\Delta)+(\epsilon_2\epsilon_\Delta))$.
  Thus, in this case we get $g_{t,c}(n,nc,b) \le g_3(n)$ where \[g_3(n) = n^{\O(1)}2^{\tw(G)}=n^{\O(1)}2^{|X_0|^{f(\lambda,\epsilon_\Delta,\epsilon_E,\epsilon_1,\epsilon_2)}}.\]

  Thus, we obtain $g_{t,c}(n,nc,b) \le \max(g_1(n,nc,b),g_2(n,nc,b),g_3(n))$.

  \paragraph*{Analysis of $g_{t,c}$.}
  Let $m(nc,b)=nc+3p(|X_0|,|X_0|)|X_0|b$ and $h(n,m)$ be the maximum of $g_{t,c}(n,nc,b)$ when $m(nc,b) \le m$.
Let us now consider any $(n,nc,b)$ such that $m(nc,b) \le m$, and bound $g_1(n,nc,b)$ and $g_2(n,nc,b)$.

\noindent\textbf{Bound on $g_1$.}\\
Notice that
\begin{align*}
m(2p(|X_0|,|X_0|)|X_0|, b-t)
&\le m(nc,b) - (3t-2)p(|X_0|,|X_0|)|X_0|\\
&\le m(nc,b) - 2t\,p(|X_0|,|X_0|)|X_0|, \mbox{ and}\\
m(nc-1,b) &\le m(nc,b) - 1,
\end{align*}
implying
  \begin{align*}
g_1(n,nc,b)
&\le n^{\O(1)} + h\bigl(n, m-2t\,p(|X_0|,|X_0|)|X_0|\bigr) + h(n,m-1) \\
&\le n^{\O(1)} + (c+1)h\bigl(n, m-2t\,p(|X_0|,|X_0|)|X_0|\bigr) + h(n,m-1).
\end{align*}

\medskip

\noindent\textbf{Bound on $g_2$.}\\
Notice that
\begin{align*}
m(2p(|X_0|,|X_0|)|X_0|, b-1)
&\le m(nc,b) - p(|X_0|,|X_0|)|X_0|, \mbox{ and}\\
m(2p(|X_0|,|X_0|)|X_0|, b-c)
&\le m(nc,b) - (3c-2)p(|X_0|,|X_0|)|X_0| \\
&\le m(nc,b) - 2tp(|X_0|,|X_0|)|X_0|,
\end{align*}
implying
\begin{align*}
g_2(n,nc,b)
&\le n^{\O(1)} + (c+1)h\bigl(n, m-2tp(|X_0|,|X_0|)|X_0|\bigr) + h\bigl(n, m-p(|X_0|,|X_0|)|X_0|\bigr)
    \\
&\le n^{\O(1)} + (c+1)h\bigl(n, m-2tp(|X_0|,|X_0|)|X_0|\bigr) + h\bigl(n, m-1\bigr).
\end{align*}

Let us now bound $h$ by taking the maximum over the three cases, which simplifies to the maximum of two values, 
since the upper bounds on $g_1(n,nc,b)$ and $g_2(n,nc,b)$ are the same. In fact,
\begin{eqnarray*}
    h(n,m) & \le& max(n^{\O(1)}+(c+1)h(n,m-2tp(|X_0|,|X_0|)|X_0|)+h(n,m-1),g_3(n)) \\
    & \le & g_3(n)+n^{\O(1)}+(c+1)h(n,m-2tp(|X_0|,|X_0|)|X_0|)+h(n,m-1). 
  \end{eqnarray*}
This leads to $h(n,m) = \O(mg_3(n) + m(c+1)h(n,m-2tp(|X_0|,|X_0|)|X_0|)$, and then to
$h(n,m) = g_3(n){(mc)}^{\O(\frac{m}{tp(|X_0|,|X_0|)|X_0|})}$. 
 
This implies that for any $(n,nc,b)$, we get (where $m=m(nc,b) = \O(p(|X_0|,|X_0|)|X_0|^2)$)
  \begin{eqnarray*}
    g_{t,c}(n,nc,b) & \le & h(n,m) \\
        & \le & h(n,\O(p(|X_0|,|X_0|)|X_0|^2)) \\
    & \le & g_3(n)|X_0|^{\O(\frac{|X_0|}{t})} \\
    & = & g_3(n)2^{\O(|X_0|^{(1-\epsilon_1)}\log(|X_0|))}\\
    & = & n^{\O(1)}2^{\O(\log(|X_0|)|X_0|^{max(f(\lambda,\epsilon_r,\epsilon_\Delta,\epsilon_E,\epsilon_1,\epsilon_2),1-\epsilon_1)})}.
  \end{eqnarray*}

Thus, given the initial instance $(G_0,X_0,k_0)$, the total running time is in
  \[
  |\I_{\epsilon_1}|g_{t,c}(n,nc,b)= n^{\O(1)}2^{\O(\log(|X_0|)|X_0|^{max(f(\lambda,\epsilon_r,\epsilon_\Delta,\epsilon_E,\epsilon_1,\epsilon_2),1-\epsilon_1)})}.
  \]
It remains to choose $\epsilon_1$ and $\epsilon_2$ that minimizes $\max(f(\lambda,\epsilon_r,\epsilon_\Delta,\epsilon_E,\epsilon_1,\epsilon_2),1-\epsilon_1)$.
By definition of $f$, it corresponds to minimizing
\[
\max \left\{
\frac{\epsilon_2}{\epsilon_r},\;
(1+(\lambda\epsilon_1+(\lambda+3)\epsilon_2)+\epsilon_1)\epsilon_E,\;
(1+(\lambda\epsilon_1+(\lambda+3)\epsilon_2))(1-\epsilon_\Delta)+(\epsilon_2\epsilon_\Delta),\;
1-\epsilon_1
\right\}.
\]
Setting $\epsilon_2=\epsilon_1$, and bounding $(2\lambda+3)\epsilon_1$ by $2(\lambda+2)\epsilon_1$, we get that, for $a = (2\lambda + 5)\epsilon_E$ and $b = 2(\lambda + 2)(1 - \epsilon_{\Delta}) + \epsilon_{\Delta}$, it is enough to minimize $\max(u_1(\epsilon_1),u_2(\epsilon_1),u_3(\epsilon_1),u_4(\epsilon_1))$
where $u_1(\epsilon_1) = \frac{\epsilon_1}{\epsilon_r}$, $u_2(\epsilon_1) = \epsilon_E + a\epsilon_1$, $u_3(\epsilon_1) = (1-\epsilon_\Delta) + b\epsilon_1$, 
$u_4(\epsilon_1) = 1 - \epsilon_1$.
Thus, the first three functions are increasing in $\epsilon_1$, and the last one is decreasing.
For $i \in \{1,2,3\}$, let $\epsilon^\star_i$ be such that $u_i(\epsilon^\star_i) = u_4(\epsilon^\star_i)$.
Notice that $\epsilon^\star_1 = \frac{\epsilon_r}{\epsilon_r+1}$, $\epsilon^\star_2 = \frac{1-\epsilon_E}{a+1} = \frac{1-\epsilon_E}{(2\lambda+5)\epsilon_E+1}$, and $\epsilon^\star_3 = \frac{\epsilon_{\Delta}}{b+1} = \frac{\epsilon_{\Delta}}{2(\lambda+2)(1-\epsilon_{\Delta})+\epsilon_{\Delta}+1}$.
Now, observe that $\min \max \{u_1,u_2,u_3,u_4\}$ is reached for $\epsilon^\star = \min(\epsilon^\star_1,\epsilon^\star_2,\epsilon^\star_3)$ and has value $1 - \epsilon^\star$.
\end{proof}

Let us now apply the previous theorem on specific graph classes.
Notice that $s$-string and pseudo-disk graphs are both subclasses of string graphs, and thus both have $(\frac{1}{4},\frac{1}{2},\frac{1}{6})$-modulator bounded treewidth by \autoref{thm:tw-string-XR}. Together with \autoref{lemma:boundedNC} and \autoref{lemma:boundedNC-pseudo}, it follows that $s$-string graph is a $(p,\frac14,\frac12,\frac16)$ PATH-Mod compliant class, where $p(x,y)=\O(x^4y^3)$. Likewise, pseudo-disk graph is a $(p',\frac14,\frac12,\frac16)$ PATH-Mod compliant class, where $p'(x,y)=\O(x^7y^3)$. This leads to the following.

\begin{corollary}\label{cor:path}
$\VC/\disto{paths}$ can be solved in time
\begin{itemize}
    \item $n^{\O(1)}2^{\log(|X|)|X|^{1-1/67}}$ in $s$-string graphs,
    \item $n^{\O(1)}2^{\log(|X|)|X|^{1-1/57}}$ in pseudo-disk graphs.
\end{itemize}
\end{corollary}
\section{Subexponential algorithms and dichotomy for \texorpdfstring{$\ASQGM(\omega)$}{ASQGM(omega)} classes}
\label{sec:asqgm}

In the following, we sometimes refer to (minor)-hitting$_{\F}/\disto{\C}$ for any statement that holds for both hitting$_{\F}/\disto{\C}$ and minor-hitting$_{\F}/\disto{\C}$. Given a value $w$, we define an \autoref{def:algoASQGM} $A_w((G,X,k))$ that decides an instance of (minor)-hitting$_{\F}/\disto{\C}$.
We assume that $B$ is an algorithm that decides (minor)-hitting$_{\F}$ parameterized by $\tw$.
In the following $n_{\F}$ denotes the minimum number of vertices, among the graphs in $\F$.    

\begin{algorithm}[!h]
\caption{$A_w((G,X,k))$}
\begin{algorithmic}[1]

    \State Try to find a clique $K$ of size $w$ in $G[X]$ in $|X|^w n^{\O(1)}$.
    \State  If $|K| < w$ then the algorithm stops and return $B(G,k)$.  //DP base case 
    \If{$k \le w-n_{\F}$}
    \State return no 
    \Else
    \State return $\bigvee_{U \subseteq K, |U|= n_{\F}-1}A_w((G\setminus (K \setminus U), X \setminus (K \setminus U) ,k-(w-n_{\F}+1))$
    \EndIf
\end{algorithmic}
\label{def:algoASQGM}
\end{algorithm}

\begin{theorem}
    \label{thm:subexpASQGM}
     Let $c$ be an integer and be $\mathcal{C}$ a hereditary graph class such that $\tw(\C)\le c$.
    Let $\mathcal{G}$ a hereditary graph class that has the $\ASQGM(\omega)$ property, i.e., there exists $\lambda > 0$ and $\lambda'>0 $ such that for all $G \in \mathcal G$, we have $\tw(G)=\O(\omega(G)^\lambda \cdot \grid(G)^{2-\lambda'})$.
    
    If (minor)-hitting$_{\F}$ can be solved in $n^{\O(1)}2^{\O(\tw(G)\log(\tw(G)))}$, then, there exists $\epsilon(\F,c,\lambda{,\lambda'}) > 0$ such that, when restricted to input graphs in $\mathcal{G}$, (minor)-hitting$_{\F}$ can be solved in time $n^{\O(1)}2^{|X|^{1-\epsilon(F,c,\lambda{,\lambda'})}}$.

More precisely, 
the running time is 
$n^{\O(1)}    2^{\O(|X|^{1-\frac{\lambda'}{2(\lambda+1)}}(c^{\lambda+2}+n_{\F})\log|X|)}$. 
\end{theorem}

\begin{proof}
  
  Let us now start with three claims on cliques.

    \begin{claim}\label{claim:distoBidim}
    Let $c'$ be an integer and $\mathcal{\C'}$ be a graph class such that $\tw(\C')\le c'$.
    Then, for any graph $G$, $\disto{\C'}(G) \ge \lfloor\grid(G) / (c'+1)\rfloor^2$.
  \end{claim}
  \begin{proof}
Consider a model of a $g \times g$ grid of $G$ where $g=\grid(G)$.
    As $\tw(\C') \le c'$, the $(c'+1) \times (c'+1)$ grid does not belong to $\C'$. As we can find $\lfloor\frac{g}{c'+1}\rfloor^2$ disjoint models of such a grid in the grid in $G$, we get that $\disto{\C'}(G) \ge \lfloor\frac{g}{c'+1}\rfloor^2$.
  \end{proof}

The two following claims are immediate.
  \begin{claim}\label{claim:sol}
    Let $(G,X,k)$ be an input of (minor)-hitting$_{\F}$.
    Let $K$ be clique of $G$, and $S$ be a solution to $(G,X,k)$. Then, $|K \setminus S| < n_{\F}$.
  \end{claim}

  \begin{claim}\label{claim:clique}
    Let $(G,X,k)$ be an input of (minor)-hitting$_{\F}/\disto{\C}$, and $c$ be an integer such that $\tw(G\setminus X) \le c$. Then, $\omega(G) \le \omega(G[X]) + c+1$. 
  \end{claim}

  The next claim follows from the fact that in algorithm $A_w$, the part outside $X$ never changes.
\begin{claim}\label{claim:valid}
  Let $(G,X,k)$ be an input of $A_w$. Any input $(G',X',k')$ given in the recursive calls in the final line is a valid input of (minor)-hitting$_{\F}/\disto{\C}$, implying in particular that $G' \setminus X' \in \C$, and moreover, $G' \in \G$ as $\G$ is hereditary.
  \end{claim}

    \begin{claim}\label{claim:outputtw}
Let $(G,X,k)$ be an input of $A_w$ (with $G\in \G$ and $G\setminus X \in \cC$) which falls into the base case treated by algorithm $B$. Then, $\tw(G)=\O((w+c)^\lambda (c \sqrt{|X|})^{2-\lambda'})$.
  \end{claim}
  \begin{proof}
By definition of $A_w$, $\omega(G[X]) \le w$. By \autoref{claim:valid}, $G \setminus X \in \C$, implying by \autoref{claim:clique} that $\omega(G) \le w+c+1$.
    As $G\setminus X \in \C$, we also get that $|X| \ge \disto{\C}(G)$ and thus \autoref{claim:distoBidim}
    implies that $|X| \ge \lfloor\frac{\grid(G)}{c+1}\rfloor^2$, implying $\grid(G) = \O(c\sqrt{|X|})$.
    Now, as $G \in \G$, it has the $\ASQGM(\omega)$ property, implying that $\tw(G)=\O((\omega(G)^\lambda \cdot \grid(G)^{2-\lambda'})=\O((w+c)^\lambda (c \sqrt{|X|})^{2-\lambda'})$
     by \autoref{claim:clique}.
  \end{proof}

  \begin{claim}\label{claim:safe}
$A_w$ correctly decides $(G,X,k)$.
  \end{claim}
  \begin{proof}
    It is immediate that if one of the enumerated $U$ is such that $A_w((G\setminus (K \setminus U), X \setminus (K \setminus U) ,k-(w-n_{\F}+1))$ returns yes,
    then $(G,X,k)$ is a yes-instance.
    The other direction follows from \autoref{claim:sol}.
  \end{proof}

Now, given $\epsilon > 0$ and an instance $(G,X,k)$, we define $w=|X|^{\epsilon}$, and run $A_w(G,X,k)$.
  Let us now analyze the running time of $A_w$.
  Let $f_w(n,x)$ be the worst case running time of a call to $A_w(G,X,k)$ for a graph $G$ and $X$ with $|V(G)| \le n$ and $|X| \le x$.
  Let $g(n,tw) =n^{\O(1)}2^{tw\log(tw)}$ be the worst case for $B(G,k)$ when $|V(G)| \le n$ and $\tw(G) \le tw$.

  If we fall into the base case then $f_w(n,x)=f^1_w(n,x)$ with $f^1_w(n,x)=g(n,\O((w+c)^\lambda (c \sqrt{|X|})^{2-\lambda'})$ by \autoref{claim:outputtw}.
  Otherwise, we get $f_w(n,x) \le f^2_w(n,x)$ where $f^2_w(n,x)= n^{\O(1)}x^w+w^{n_{\F}}f_w(n,x-(w-n_{\F}))$.
  As $f_w(n,x) \le f^1_w(n,x)+f^2_w(n,x)$, we get
  \begin{eqnarray*}
      f_w(n,x) &\le & n^{\O(1)}x^w+f^1_w(n,x)+w^{n_{\F}}f_w(n,x-(w-n_{\F})).
  \end{eqnarray*}
  This implies
  \begin{eqnarray*}
      f_w(n,x) &\le & (n^{\O(1)}x^w+f^1_w(n,x))w^{\frac{xn_{\F}}{w-n_{\F}}}.
  \end{eqnarray*}
  Now, observe that
  \begin{eqnarray*}
     n^{\O(1)}x^w+f^1_w(n,x)  &\le & n^{\O(1)}(2^{\O(x^\epsilon \log x)}+2^{\O(\lambda c^{\lambda+2}x^{\lambda\epsilon+1-\frac{\lambda'}{2}}\log|X|)}) \\
     &= & n^{\O(1)}2^{\O(\lambda c^{\lambda+2}x^{\lambda\epsilon+1-\frac{\lambda'}{2}}\log|X|)},
  \end{eqnarray*}
This leads to
\begin{eqnarray*}
      f_w(n,x) &\le & n^{\O(1)}2^{\O(\lambda c^{\lambda+2}x^{\lambda\epsilon+1-\frac{\lambda'}{2}}\log|X|)}  w^{\frac{xn_{\F}}{w-n_{\F}}} \\
      &\le & n^{\O(1)}2^{\O(\log|X|(\lambda\epsilon+1-\frac{\lambda'}{2}+\epsilon x^{1-\epsilon}n_{\F}))}
\end{eqnarray*}

Now, we chose $\epsilon$ such that $1-\epsilon=\lambda\epsilon+1-\frac{\lambda'}{2}$, leading to $\epsilon=\frac{\lambda'}{2(\lambda+1)}$ and to the claimed running time.
\end{proof}

Notice that, in particular, as soon as $\F$ contains at least one planar graph, minor-hitting$_{\F}$ can be solved in $n^{\O(1)}2^{\tw(G)\log(\tw(G))}$~\cite{baste2023hitting}, 
and the same running time holds for the problem hitting$_{P_x}$ where $P_x$ is the path on $x$ vertices~\cite{cygan2017hitting}.

Recall that as any minor-closed family $\mathcal{C}$ of unbounded treewidth contains all planar graph, and as for any $\F$ containing connected graphs (of at least two vertices) and at least one planar one, (minor)-hitting$_{\F}$ remains \NPH when restricted to planar graphs~\cite{yannakakis1978node}. In that case  $\disto{\mathcal{C}}=0$, so the parameterized problem is not XP (see \autoref{obs:lbASQGM}).

\autoref{thm:subexpASQGM} and \autoref{obs:lbASQGM} leads to the following dichotomy.

\begin{corollary}
  \label{cor:dicho}
    Let $\F$ be a finite set of connected graphs (of at least two vertices)
    such that (minor)-hitting$_{\F}$ can be solved in $n^{\O(1)}2^{\O(\tw(G)\log(\tw(G)))}$ time,
    and such that at least one of them is planar.
  Let $\mathcal{C}$ be a minor-closed family, and $\mathcal{G}$ a hereditary graph class containing planar graphs that has the $\ASQGM(\omega)$ property.
  Then, unless $P=NP$, (minor)-hitting$_{\F}/\disto{\C}$  admits a subexponential algorithm iff $\tw(\C)$ is bounded.
\end{corollary}
\section{Proving that s-string and pseudo-disks graphs are PATH-Mod compliant classes and that string graphs is a TD-Mod compliant class}\label{sec:graph}

The classes of string graphs, $s$-string graphs and pseudo-disk graphs are hereditary, so \Cref{thm:ss-pd-type12} follows by showing that the latter two classes have $p$-bounded neighborhood complexity for some polynomial $p$ (see \Cref{ssec:pbnc}), and by showing that string graphs (a super-class of $s$-string graphs and pseudo-disk graphs) have modulator bounded treewidth (see \Cref{ssec:mbt}) and
bounded large neighborhood components (see \Cref{ssec:blnc}).

\subsection{String graphs have modulator bounded treewidth}\label{ssec:mbt}

\begin{theorem}\label{thm:tw-string-XR}
String graphs have $(\frac{1}{4},\frac{1}{2},\frac{1}{6})$-modulator bounded treewidth.
\end{theorem}

The proof requires the following two properties of string graphs.
\begin{theorem}[Combining \cite{lee2016separators} and \cite{DVORAK2019137}] \label{th:sep-m-tw}
  Any $m$-edge string graph has treewidth $\O(\sqrt{m})$.
\end{theorem}
\begin{lemma}\label{lem:nb-induced-XRX}
  Consider a string graph $G$, and a partition of its vertices into two sets $X,R$ such that $R$ is an independent set, and such that $\Delta_X \le |X|^{1/4}$. Then, there are $\O(|X|^{5/3}\Delta_X^{1/3})$ non-adjacent pairs $u,v\in X$ with a common neighbor $r\in R$.
\end{lemma}
\begin{proof}
Let us bound the size of the set $M\subseteq \binom{X}{2}$ of the pairs of non-adjacent vertices of $X$ having a common neighbor (at least) in $R$.

Suppose $S$ is a subset of $X$ of size $k$ chosen uniformly at random, for $k=\left\lfloor(|X|/\Delta_X)^{1/3}\right\rfloor$. Trivially, 
$$
k^4\Delta_X / |X| \le  k.
$$
Moreover, assuming $|X|$ is larger than a constant, we have
$$
\frac{\Delta_X(\Delta_X+k)}{|X|} <  \frac12 .
$$
Indeed, $\Delta_X^2/|X| \le 1/|X|^{1/2} \le \frac14$, and
$\Delta_X k/|X|\le (\Delta_X/|X|)^{2/3} \le |X|^{(-3/4)\cdot(2/3)}\le \frac14$.

Let $Span(S)\subseteq M$ be the subset of pairs with both vertices in $S$.
Note that $$|M| \binom{|X|-2}{k-2} = \sum_{i=1}^{|M|} \binom{|X|-2}{k-2} = \sum_{\text{$k$-vertex subsets $S$}} |Span(S)| = \binom{|X|}{k} \mathbb{E}[|Span(S)|],$$ 
where the second equality follows by double counting the pairs $(\{u,v\},S)$ such that $\{u,v\} \in M$ and $S$ is a $k$-vertex subset containing both $u$ and $v$. On the one hand, for each pair $\{u,v\}$, there are $\binom{|X|-2}{k-2}$ such subsets $S$. On the other hand, for each $k$-vertex subset $S$, there are $|Span(S)|$ such pairs $\{u,v\}$.

Hence, $|M| = \O(|X|^2 / k^2) \mathbb{E}[|Span(S)|]$, and we just need to prove that $\mathbb{E}[|Span(S)|] = \O(k)$.
A pair $u,v \in X^2$, is {\em selected by $S$} if there is a vertex $r\in R$ (at least), such that $N_X(r)\cap S = \{u,v\}$. Let $Sel(S) \subseteq Span(S)$ be the set of pairs selected by $S$.

\begin{claim}
For any pair $\{u,v\}\in M$, $Pr[\{u,v\} \in Sel(S)] > \frac12 Pr[\{u,v\} \in Span(S)]$. 
\end{claim}
\begin{proof}
    For any pair $\{u,v\}\in M$, let us denote $r_{u,v}$ one of the vertices of $R$ such that $\{u,v\}\in N_X(r_{u,v})$, and note the following:
$$
Pr[\{u,v\} \in Sel(S)]  \ge  Pr[\{u,v\}\in Span(S)]\times Pr[N_X(r_{u,v}) \cap S = \{u,v\}\ |\ \{u,v\}\in Span(S)].
$$
    Let us now bound the latter probability. We have
    \begin{eqnarray*}
    Pr[N_X(r_{u,v}) \cap S = \{u,v\}\ |\ \{u,v\}\in Span(S)] & = & \binom{|X|-d_X(r_{u,v})}{k-2} / \binom{|X|-2}{k-2}\\
    & \ge &  \frac{(|X|-\Delta_X)!\ (|X|-k)!}{(|X|-\Delta_X -k+2)!\ (|X|-2)!}\\ 
    & \ge & \frac{(|X|-k)\ldots(|X|-\Delta_X -k+3)}{(|X|-2)\ldots(|X|-\Delta_X+1)}\\ 
    & > & \left(\frac{|X|-\Delta_X-k}{|X|}\right)^{\Delta_X-2}\\
    & > & \left(1 - \frac{\Delta_X+k}{|X|}\right)^{\Delta_X}\\
    & > & \left(1 - \frac{\Delta_X(\Delta_X+k)}{|X|}\right)\\
    & > & \frac12.
    \end{eqnarray*} 
\end{proof}

The claim above implies that $\mathbb{E}[|Span(S)|] < 2 \mathbb{E}[|Sel(S)|]$, and thus we just need to prove that $\mathbb{E}[|Sel(S)|] = \O(k)$.

\begin{remark}\label{rmk:big-sel}
    By definition of $M$, for any set $S\subseteq X$, we have that $|Sel(S)| <k^2$.
\end{remark}

This bound is too large, but under some conditions, the set $Sel(S)$ can be shown much smaller.
Let $\I$ denote the $k$-vertex subsets of $X$ that induce an independent set.

\begin{claim}\label{claim-sel-I}
    If $S \in \I$, then $|Sel(S)| < 3k$.
\end{claim}
\begin{proof}
    For each pair $\{u,v\}\in Sel(S)$, let us denote $r'_{u,v}$ one of the vertices of $R$ such that $N_X(r'_{u,v})\cap S = \{u,v\}$. Note that all these vertices are distinct, $\{u,v\}\neq \{x,y\} \Longrightarrow r'_{u,v}\neq r'_{x,y}$.  
    Consider the string graph $G_S$ induced by $S$ and by the vertices $r'_{u,v}$. Clearly, $G_S$ is a $1$-subdivision of a simple $k$-vertex graph. Hence, we have that $G_S$ is a 1-subdivision of a simple planar graph with $k$ vertices~\cite[Section 2]{ehrlich1976intersection} and at most $3k-6$ edges, implying that $Sel(S)$ has size at most $3k-6$.
\end{proof}

Let us now show that a randomly chosen set $S$ is likely to be an independent set.
\begin{remark}\label{rem:pr-S-indep}
  $Pr[S \notin \I] \leq \mathbb{E}[|E(G[S])|] \le \Delta_X |X| \frac{k(k-1)}{|X|(|X|-1)} < k^2\Delta_X / |X|$.
\end{remark}

We can now conclude
\begin{eqnarray*}
    \mathbb{E}[|Sel(S)|] & = &  Pr[S \in \I]\cdot\mathbb{E}[|Sel(S)|\ |\ S \in \I] \\
    & & + Pr[S \notin \I]\cdot\mathbb{E}[|Sel(S)|\ |\ S \notin \I] \\
    & < & 3k + k^2\Delta_X / |X| \cdot k^2\\ 
    & < & 3k + k^4\Delta_X / |X|\\
    & < & 4k.
\end{eqnarray*}
This concludes the proof of~\cref{lem:nb-induced-XRX}.
\end{proof}

\begin{proof}[Proof of \Cref{thm:tw-string-XR}]
Consider a string graph $G$, and a partition of its vertices into two sets $X,R$ such that for every connected component $C$ of $G[R]$, $|N_X(C)|\le \Delta$, for some $\Delta \le |X|^{1/4}$. We have to show that $\tw(G) = \O(m_X^{1/2} + \tw(G[R]) +|X|^{5/6}\Delta^{1/6})$, where $m_X$ is the number of edges of $G[X]$.

Let $G'_X$ be the graph obtained from $G[X]$ by turning $N_X(C)$ into complete graphs, for every connected component $C$ of $G[R]$.
\begin{claim}
$G'_X$ is a string graph.
\end{claim}
This follows from the fact that the class of string graphs is stable under edge contraction, and under the addition of true- or false-twins. Start from $G$ by contracting each connected component $C$ of $G[R]$ into a single vertex $v_C$, then duplicate each $v_C$ into $|N_X(C)|$ copies, and finally contract a matching between these copies and $N_X(C)$. The obtained graph is $G'_X$, and it is thus a string graph.
\begin{claim}
$\tw(G'_X) = \O(m_X^{1/2} + |X|^{5/6}\Delta^{1/6})$.
\end{claim}
\Cref{lem:nb-induced-XRX} implies that $G'_X$ has at most $m_X + \O(|X|^{5/3}\Delta^{1/3})$ edges. Hence by~\Cref{th:sep-m-tw}, 
$$\tw(G'_X) = \O(\sqrt{m_X + \O(|X|^{5/3}\Delta^{1/3})}) = \O(m_X^{1/2} + |X|^{5/6}\Delta^{1/6}).$$

Now consider the super-graph of $G$, $G'$, that is the (non-disjoint) union of $G$ and $G'_X$, and let us show that $\tw(G') \le \tw(G'_X) + \tw(G[R])+1$, concluding our proof. Indeed, consider optimal tree-decompositions of $G'_X$, $(T_{G'_X},\beta_{G'_X})$, and of each connected component $C$ of $G[R]$, $(T_C,\beta_C)$. We construct a tree-decomposition of $G'$, $(T',\beta')$, as follows. The tree $T'$ is obtained by linking $T_{G'_X}$ with $T_C$ through an edge $xc$, where $x\in V(T_{G'_X})$ is a node such that $N_X(C)\subseteq \beta_{G'_X}(x)$ (which exists since $N_X(C)$ induces a clique in $G'_X$), and where $c$ is any node of $T_C$. The mapping $\beta'$ is obtained from $\beta_{G'_X}$ and the mappings $\beta_{C}$ by adding $N_X(C)$ in the image of each nodes of $T_C$. The sizes of these bags increase by at most $\tw(G'_X)+1$.
It is routine to check that the obtained $(T',\beta')$ forms the expected tree-decomposition of $G'$, and thus of $G$.
\end{proof}

\subsection{Polynomial bounded neighborhood complexity}\label{ssec:pbnc}

For the class of pseudo-disk graphs, their $p$-bounded neighborhood complexity follows from the Lemma 5.2 of~\cite{Berthe2025}.
There, the variant of the neighborhood complexity considered, dubbed \emph{tree neighborhood complexity}, considers the size of the connected components in $G\setminus X$ as a parameter, as well as the minimal $r$ such that $G$ is $K_{r,r}$-free.
In the proof of Lemma 5.2, these parameters do not interfere, so we can state the following. 
\begin{theorem}[Lemma 5.2 of~\cite{Berthe2025}]\label{lemma:boundedNC-pseudo}
The class of pseudo-disk graphs has $p$-bounded neighborhood complexity for $p(\Delta(G[X]),d)= \O(d^3)$.
\end{theorem}

For showing the $p$-bounded neighborhood complexity of $s$-string graphs we cannot follow exactly the lines of Lemma 5.4 from~\cite{Berthe2025}, as there the mentioned parameters are used.
However our proof remains very similar and also uses the following result.

\begin{theorem}[\cite{pseudodisjoint}]\label{hyperpseudo}
    Given $\EuScript{F}$ a family of pseudo-disks and $\EuScript B$ a finite family of pseudo-disks, consider the hypergraph $H(\EuScript B,\EuScript F)$ whose vertices are the pseudo-disks in $\EuScript B$ and the edges are all subsets of~$\EuScript B$ of the form $\{D \in \EuScript B,~D \cap S \neq \emptyset\}$, with $S\in \EuScript F$.
  Then the number of edges of cardinality at most $k\geq 1$ in $H(\EuScript B,\EuScript F)$ is $\O(|\EuScript B|k^3)$.
\end{theorem}

\begin{figure}
    \centering
    \includegraphics[width=0.8\linewidth]{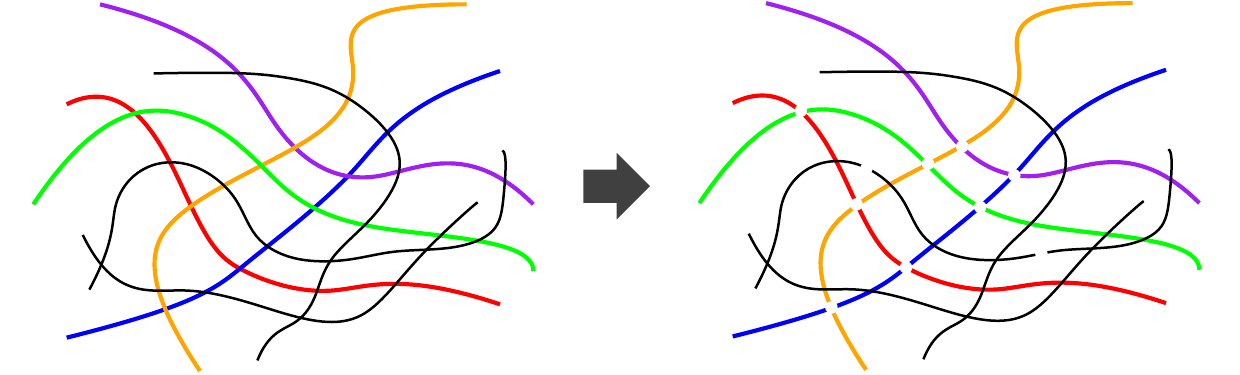}
    \caption{Cutting strings of $X$ (the colored ones) and replacing a connected component of $G[R]$ (in black) by a pseudo disk.}
    \label{fig:cutting-s-strings}
\end{figure}

\begin{theorem}
\label{lemma:boundedNC}
For any $s\ge 1$, the class of $s$-string graphs has $p$-bounded neighborhood complexity for $p(\Delta(G[X]),d) = \O(d^3 s^4 (\Delta(G[X])+1)^4)$.
\end{theorem}
\begin{proof}
Let us show that for any $s$-string graph $G$, and any set $X\subseteq V(G)$, we have that $\big|\{N_X(C) \mid C \in \mathcal{C}^{\le d}(G,X)\}\big| \le p(\Delta(G[X]),d)|X| = \O(d^3 s^4 (\Delta(G[X])+1)^4)\cdot |X|$.

Note that every string of $X$ contains at most $s\Delta(G[X])$ intersection points with another string of $X$. Cutting them at these points, each string of $X$ gives rise to $s\Delta(G[X])+1$ substrings, all these substrings being disjoint from each other (see \Cref{fig:cutting-s-strings}). Thickening them, we can assume they form a family of (disjoint) pseudo-disks $\mathcal{B}$ of size $\O(|X|(s\Delta(G[X])+1))$. Now each connected component of $\mathcal{C}^{\le d}$ intersects at most $d\times (s\Delta(G[X])+1)$ of these substrings. We can replace each connected component of $\mathcal{C}^{\le d}$ with a pseudo-disk intersecting the same substrings. Indeed, this pseudo-disk is obtained starting from the union of the strings, by iteratively cutting any closed curve contained in this set on a small part, and when reaching a point where this (connected set) does not contain any closed curve, by slightly thickening the arcs of this set (see \Cref{fig:cutting-s-strings}). The connected components being disjoint from each other, they also lead to a family of disjoint pseudo-disks $\mathcal{F}$. Consider now the hypergraph $H(\mathcal{B},\mathcal{F})$ of \Cref{hyperpseudo}. This hypergraph has no multiple edges, so its number of edges correspond to distinct neighborhoods. Hence \Cref{hyperpseudo} implies that our set of neighborhoods has size $\O(|\mathcal{B}|(d (s\Delta(G[X])+1))^3) = \O(d^3 s^4 (\Delta(G[X])+1)^4 |X|)$.
\end{proof}

\subsection{Bounded number of large neighborhood components}\label{ssec:blnc}

Here, we need the following alternative definition of treedepth. The \emph{treedepth} of a graph $G$, is the minimum number $\td$ such that $G$ admits a vertex coloring $V(G)\longrightarrow [1,\td]$ with the following property: for every connected induced subgraph $H$ of $G$, the lowest color in $H$ appears exactly once. Note that such a coloring is proper, every edge has distinct colors at its endpoints.

\begin{theorem}\label{lem:blnc}
    For any string graph $G$, any set $X\subseteq V(G)$, and any $\td\ge \td(G\setminus X)$, either $|X| = \O(\Delta_X^5)$, or $|\mathcal{C}^{\ge 6\td \Delta_X^2}| = \O_{\td}(\Delta_X^7) \cdot |X|$.
\end{theorem}
\begin{proof}
For every connected component $C$ in $\mathcal{C}^{\ge 6\td \Delta_X^2}$,
let us denote $\mathbb{S}_C$ the union of the strings in $C$. Within those, we are going to look for tripods. A \emph{tripod} is a Y-shaped region (i.e. the union of 3 curves intersecting only at one of their endpoints), such that their 3 endpoints intersect 3 strings of $X$ pairwise non-intersecting, called the leaves of the tripod, while the rest of the tripod does not intersect these leaves, and cannot intersect more than $6\td \Delta_X^2 + \Delta_X$ strings of $X$. 

\begin{claim}\label{claim-tripod}
  For every connected component $C$ in $\mathcal{C}^{\ge 6\td \Delta_X^2}$, the region $\mathbb{S}_C$ contains a tripod $\mathbb{T}_C$.
\end{claim}
\begin{proof}
Let us first consider a subset $C'\subseteq C$ such that $G[C']$ is connected and $6\td  \Delta_X^2 \le |N_X(C')| < 6\td  \Delta_X^2 + \Delta_X$. We obtain this, by starting with $C'=C$ and by doing the following iteratively: While $N_X(C')$ is too large, pick any vertex $v\in C'$ such that $G[C'] - v$ remains connected, and update $C'$ by removing vertex $v$. As $d_X(v)\le \Delta_X$, we have that we necessarily stop within the indicated range. 

Now, to construct the tripod, we start with the set $\mathbb{S}$ of all the strings of $C'$, and from a labeling $\lambda : C \longrightarrow [1,\td]$ testifying the treedepth of $G\setminus X$. 

In the following steps $C'$ will be updated, but it will remain a set of strings that are substrings of the original strings. Hence the set 
$\mathbb{S}_{C'} := \cup_{\sigma\in {C'}} \sigma$ will remain a subset of $\mathbb{S}_{C}$. The set $C'$, and thus $\mathbb{S}_{C'}$, is iteratively updated as follows: for the highest possible label $\ell \in [1,\td]$ we consider a string $\sigma^*\in C'$ labeled $\ell$ and a closed curve $\mathcal{C}$ contained in $\mathbb{S}_{C'}$ going through $\sigma^*$. We cut the string $\sigma^*$ on a small part contained in $\mathcal{C}$ where no intersection occur, and the two newly created substrings keep the same label (see \Cref{fig:update-c-sc}). Note that in the corresponding intersection graph this operation consists in splitting a vertex $v^*$ into two vertices, $v^*_1,v^*_2$, with $N(v^*)= N(v^*_1) \cup N(v^*_2)$, and that the new graph has the same treedepth, as testified by the labeling. Indeed, the only connected induced subgraphs where the lowest color may not be unique now, are those having both new vertices, and all the other strings having higher labels. This would imply the existence of a closed curve $\mathcal{C}'$ whose label of $\sigma^*$, $\ell$, was the lowest, contradicting the definition of $\mathcal{C}$.
When this process terminates, $\mathbb{S}_{C'}$ does not contain closed curves, and the induced graph is hence a tree of treedepth at most $\td$.
Consider now any (non-closed) curve $\mathcal{C}'$ contained in $\mathbb{S}_{C'}$, and note that the labels of the strings of $C'$ it is going through are successively strictly decreasing and then strictly increasing. This implies that $\mathcal{C}'$ goes through at most $2\td-1$ string of $C'$. This implies that such curve goes through less than $2\td$ of the (original) strings of $C$.

\begin{figure}
    \centering
    \includegraphics[width=0.8\linewidth]{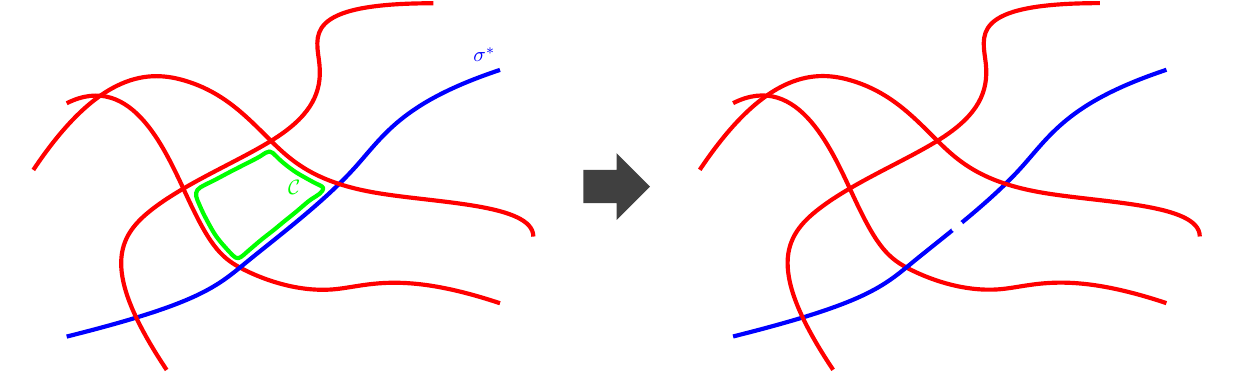}
    \caption{Cutting the string $\sigma^*$ (in blue) at a point contained in the closed curve $\mathcal{C}$ (in green), creating two new substrings with the same label.}
    \label{fig:update-c-sc}
\end{figure}

We are now going to shorten some of the strings in $C'$, until maybe erasing some of them completely (see \Cref{fig:acyclic_and_shortening}). We proceed iteratively as follows: for every endpoint $p$ of a string $\sigma \in C'$ that does not belong to another string of $C'$ we shorten $\sigma$ as much as possible. In fact we stop the shortening when the new endpoint belongs to another string of $C'$, or when it belongs to a string of $X$, that is not intersected by another string of (the current) $C'$. Strings reduced to a single point are deleted. At the end of this process every endpoint $p$ of a string $\sigma \in C'$ that does not belong to another string of $C'$ intersects a string $\sigma_x$ of $X$, and that is the unique intersection point between $\sigma_x$ and $\mathbb{S}_{C'}$. 

\begin{figure}
    \centering
    \includegraphics[width=\textwidth]{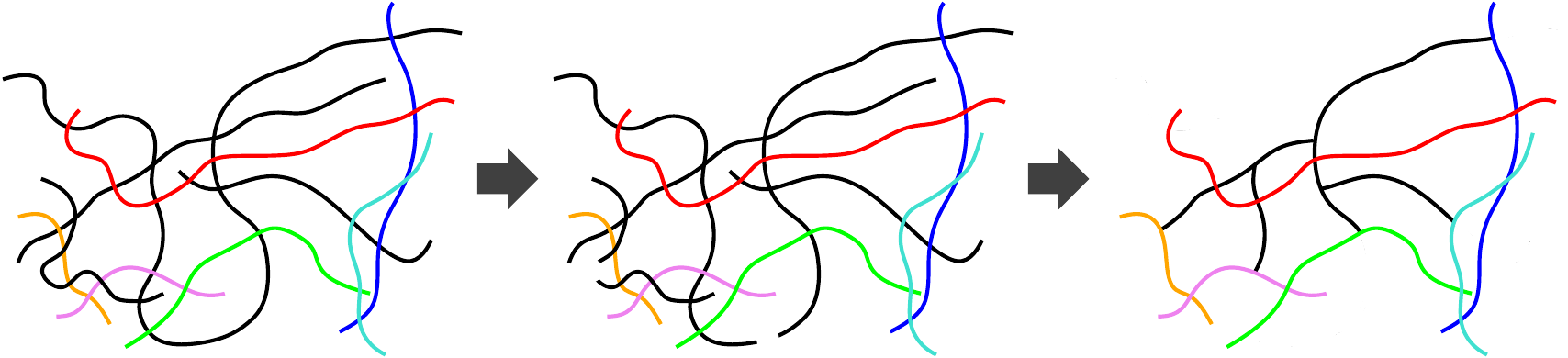}
    \caption{(left) The black strings correspond to a component $C$ of $\mathcal{C}^{\ge 6\td \Delta_X^2}$, and the colored ones to strings of $X$. (middle) The strings of $C$ are cut into a set of substrings $C'$ so that $\mathbb{S}_{C'}$ remains connected but does not contain closed curves. (right) The strings of $C'$ are shortened as much as possible while keeping the same set of intersected strings in $X$.}
    \label{fig:acyclic_and_shortening}
\end{figure}

Note that $\mathbb{S}_{C'}$ cannot be covered by $2\Delta_X+1$ curves, as otherwise it would intersect at most $(2\Delta_X+1)\cdot (2\td-1)\Delta_X$ strings of $X$, contradicting $C\in \mathcal{C}^{\ge 6\td \Delta_X^2}$.
Hence $\mathbb{S}_{C'}$ has at least $2\Delta_X+3$ endpoints, each intersecting distinct strings of $X$. Each of these strings of $X$ having at most $\Delta_X$ neighbors within those, there are three of them $\sigma_x^1,\sigma_x^2,\sigma_x^3$ that are pairwise non-intersecting. From the corresponding leaves of $\mathbb{S}_{C'}$, it is now easy to construct the tripod $\mathbb{T}_{C} \subseteq \mathbb{S}_{C'} \subseteq \mathbb{S}_{C}$ (see \Cref{fig:tripod}).
\end{proof}

\begin{figure}
    \centering
    \includegraphics[width=\textwidth]{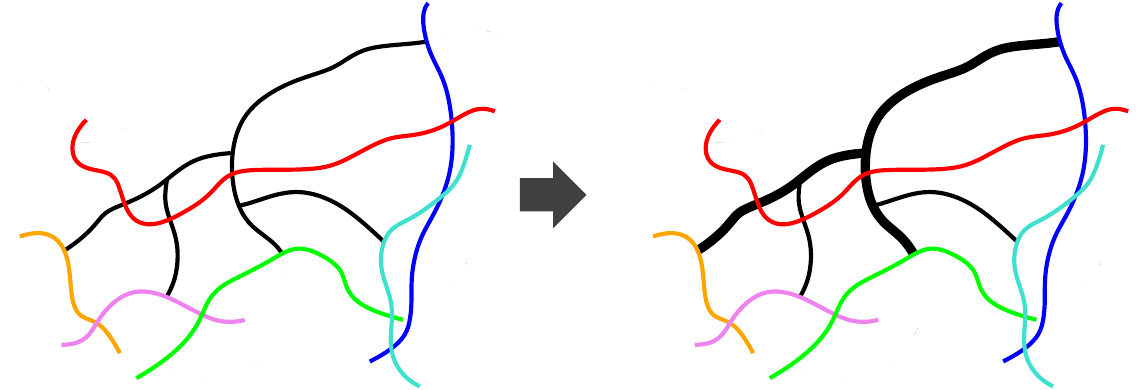}
    \caption{ The tripod obtained by considering the three non-adjacent strings at the endpoints of $\mathbb{S}_{C'}$.}
    \label{fig:tripod}
\end{figure}

Let $\mathcal{T}$ be the set of tripods, and note that $|\mathcal{T}|=|\mathcal{C}^{\ge 6\td  \Delta_X^2}|$.
To bound the size of $\mathcal{T}$, let us first bound the maximum number of tripods having their leaves in an independent subset $I$ of $X$. To this end, recall that each tripod intersects at most $\alpha := (6\td \Delta_X^2 + \Delta_X) = O_{\td}(\Delta_X^2)$ strings of $X$. Let $Span(S)\subseteq \mathcal{T}$ be the subset of tripods having their three leaves in $S$.
\begin{lemma}\label{lem:blnc-I-indep}
    For any independent subset $I$ of $X$, either 
    $|I|=\O(\Delta_X^4)$, or $|Span(I)| = \O_{\td}(\Delta_X^4 \cdot |I|)$
\end{lemma}
\begin{proof}
Consider an independent subset $I$ of $X$ such that $|I|=\Omega(\Delta_X^4)$.
This implies $\alpha(\alpha+k)/|I| < 1/2$, as $\alpha^2/|I| = o(1)$ and
$\alpha k/|I|\le \frac14$.

Now, let $S$ be a subset of $I$ of size $k$ chosen uniformly at random, for $k= \lfloor|I|/ 4\alpha\rfloor$. 
Note that $$|Span(I)| \binom{|I|-3}{k-3} = \sum_{i=1}^{|Span(I)|} \binom{|I|-3}{k-3} = \sum_{\text{$k$-vertex subsets $S$}} |Span(S)| = \binom{|I|}{k} \mathbb{E}[|Span(S)|].$$

Hence, $|Span(I)| = \O(|I|^3 / k^3) \mathbb{E}[|Span(S)|]$, and we just need to prove that $\mathbb{E}[|Span(S)|] = \O(k)$.
A tripod $\mathbb{T}_C\in Span(S)$  is {\em selected by $S$} if it does not intersect more elements of $S$, besides its leaves. Let $Sel(S) \subseteq Span(S)$ be the set of selected tripods.

\begin{claim}
For any tripod $\mathbb{T}_C$, $Pr[\mathbb{T}_C \in Sel(S)] > \frac12 Pr[\mathbb{T}_C \in Span(S)]$.
\end{claim}
\begin{proof}
As $Pr[\mathbb{T}_C \in Sel(S)] = Pr[\mathbb{T}_C\in Span(S)]\times Pr[\mathbb{T}_C\in Sel(S)\ |\ \mathbb{T}_C\in Span(S)]$, we bound the latter probability.
    \begin{eqnarray*}
    Pr[\mathbb{T}_C\in Sel(S)\ |\ \mathbb{T}_C\in Span(S)] & = & \binom{|I|-d_{I}(\mathbb{T}_C)}{k-3} / \binom{|I|-3}{k-3}\\
    & \ge &  \frac{(|I|-\alpha)!\ (|I|-k)!}{(|I|-\alpha -k+3)!\ (|I|-3)!}\\
    & \ge & \frac{(|I|-k)\ldots(|I|-\alpha -k+4)}{(|I|-3)\ldots(|I|-\alpha+1)}\\ 
    & > & \left(\frac{|I|-\alpha-k}{|I|}\right)^{\alpha-3}\\
    & > & \left(1 - \frac{\alpha+k}{|I|}\right)^{\alpha}\\
    & > & \left(1 - \frac{\alpha(\alpha+k)}{|I|}\right)\\
    & > & \frac12
    \end{eqnarray*} 
\end{proof}

The claim above implies that $\mathbb{E}[|Span(S)|] < 2 \mathbb{E}[|Sel(S)|]$, and thus we just need to prove that $\mathbb{E}[|Sel(S)|] = \O(k)$.

\begin{claim}\label{claim-sel-S}
    For any chosen set $S$, $|Sel(S)| \leq 2k$. Hence, $\mathbb{E}[|Sel(S)|] = \O(k)$.
\end{claim}
\begin{proof}
    Consider the bipartite planar graph obtained by placing vertices at the center of the tripods, and anywhere on each string of $S$, and then for each leg of the tripods, tracing an edge from the center of the tripod to the vertex on its leaf (see~\Cref{fig:tripods_2_bipartite}).
    The fact that the string of $S$ are pairwise disjoint, that the tripods are pairwise disjoint, and that among $S$ the tripods only intersect their leaves, ensures the planarity of this drawing. Then~\cref{lem:planar-linear-size} below allows us to conclude.
\end{proof}
\begin{figure}
    \centering
    \includegraphics[width=0.8\textwidth]{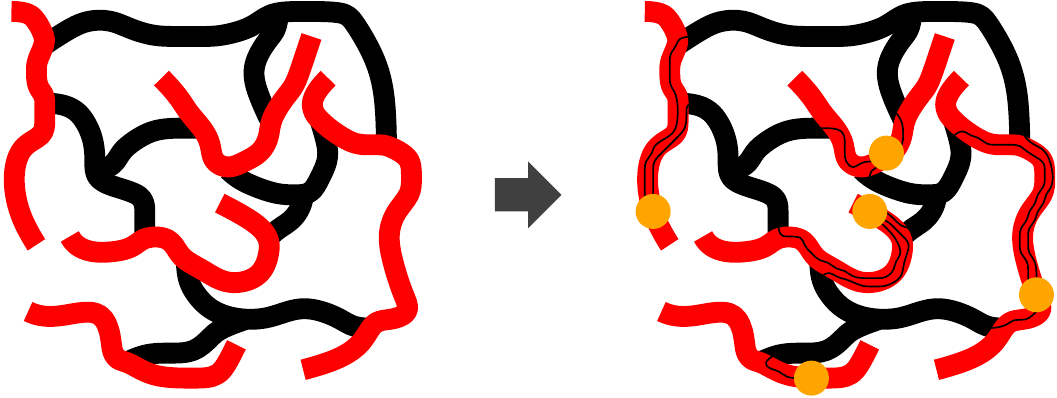}
    \caption{The bipartite graph obtained from a set of non-crossing tripods and strings.}
    \label{fig:tripods_2_bipartite}
\end{figure}

\begin{lemma}[{see \cite[Lemma~9.24]{Cygan2015Book}}]
\label{lem:planar-linear-size}
    For any bipartite planar graph $G=(A,B,E)$, if $d(v)\ge 3$ for every $v\in A$, then we have that $|A| \leq 2|B|$.
\end{lemma}
This concludes the proof of \Cref{lem:blnc-I-indep}.
\end{proof}

We are now going to bound the size of $\mathcal{T}$ through another randomized procedure. This procedure outputs a subset $I\subseteq X$ with interesting properties.
\begin{lemma}\label{lem:good_I-indep}
    There exists a probability distribution over the independent subsets $I$ of $X$ such that:
    \begin{itemize}
        \item the size of $I$ is at least $|X|/(\Delta_X+1)$,
        \item for every $v \in X$, $Pr(v \in I) \geq 1/(d_X(v)+1) \geq 1/(\Delta_X+1)$,
        \item for every tripod $\mathbb{T}_C \in \mathcal{T}$, $Pr(\mathbb{T}_C \in Span(I)) \geq 1/[(3\Delta_X+3)(3\Delta_X+2)(3\Delta_X+1)]$.
    \end{itemize}
    Therefore, $\mathbb{E}(|Span(I)|) = \Omega(|\mathcal{T}|/\Delta_X^3)$.
\end{lemma}
\begin{proof}

Consider the random process where we sort the vertices in $X$ uniformly at random, and then we construct $I$ using a greedy algorithm. Each time we select a vertex $v$, we remove all its neighbors from further consideration. Therefore, in each step we remove at most $d_X(v) + 1 \leq \Delta_X + 1$ vertices, which proves the first item.

Now, observe that a vertex $v \in X$ is selected whenever $v$ goes before $N_X(v)$. However, since the order of the $|X|$ vertices is uniform, the induced relative order on any subset is also uniform. In particular, if we focus on the subset $N_X[v]$, only the relative order of its elements matters, not their positions among the other vertices. Besides, all $(d_X(v)+1)!$ possible relative orderings are equally likely, and in exactly $d_X(v)!$ of them the vertex $v$ appears first. Therefore, we get $Pr(v \in I) \geq 1/(d_X(v)+1) \geq 1/(\Delta_X+1)$, which proves the second item.

Later, consider any tripod $\mathbb{T}_C$, and let $u,v,w \in X$ be its three leaves. Recall that $\mathbb{T}_C \in Span(I)$ if and only if $u,v,w \in I$. Denote $N_X = N_X[u] \cup N_X[v] \cup N_X[w]$ and observe that $|N_X| \leq 3\Delta_X+3$. To compute the desired probability, we apply Laplace's rule. Although there are $|X|!$ possible permutations of $X$, we will count them in a different way. First, we fix the relative order of the elements in $N_X$, which can be done in $|N_X|!$ ways. Next, we choose the positions that these elements occupy in the full ordering of $X$, giving $\binom{|X|}{|N_X|}$ possibilities. Finally, we order the $|X|-|N_X|$ remaining elements in the remaining positions, having $(|X|-|N_X|)!$ possible choices.

For the number of favorable cases, we proceed a bit differently. Instead of counting all permutations where $u$ goes before $N_X(u)$, $v$ goes before $N_X(v)$, and $w$ goes before $N_X(w)$, we will count only those where $u$ goes first, $v$ goes second, and $w$ goes third in the relative order of $N_X$. This will give us a lower bound. Then, the relative order of the $|N_X|-3$ remaining elements of $N_X$, can be done in $(|N_X|-3)!$ ways. Then, we choose the positions that the $|N_X|$ elements occupy in the full ordering of $X$, giving $\binom{|X|}{|N_X|}$ possibilities. Finally, we order the $|X|-|N_X|$ remaining elements in the remaining positions, having $(|X|-|N_X|)!$ possible choices.

Therefore, our desired probability can be lower bounded as follows:
    \begin{eqnarray*}
    Pr(u,v,w \in I) & \geq & \frac{(|N_X|-3)!\binom{|X|}{|N_X|}(|X|-|N_X|)!}{|N_X|!\binom{|X|}{|N_X|}(|X|-|N_X|)!}\\
    & = & \frac{1}{|N_X|(|N_X|-1)(|N_X|-2)}\\
    & \geq & \frac{1}{(3\Delta_X+3)(3\Delta_X+2)(3\Delta_X+1)},
    \end{eqnarray*}
which proves the third item.

To conclude, notice that
    \begin{eqnarray*}
    \mathbb{E}(|Span(I)|) & = & \sum_{\mathbb{T}_C \in \mathcal{T}} Pr(\mathbb{T}_C \in Span(I))\\
    & \geq & \sum_{\mathbb{T}_C \in \mathcal{T}} \frac{1}{(3\Delta_X+3)(3\Delta_X+2)(3\Delta_X+1)}\\
    & = & \frac{|\mathcal{T}|}{(3\Delta_X+3)(3\Delta_X+2)(3\Delta_X+1)},
    \end{eqnarray*}
which completes the proof of \Cref{lem:good_I-indep}.
\end{proof}

To finish the proof of \cref{lem:blnc} we select an independent subset $I$ of $X$ according to \autoref{lem:good_I-indep}. Assuming $|X|$ is not $\O(\Delta_X^5)$, we have that $|I|$ is not $\O(\Delta_X^4)$. Applying \autoref{lem:blnc-I-indep}, we deduce
    \begin{eqnarray*}
    |\mathcal{C}^{\ge 6\td  \Delta_X^2}| & = & |\mathcal{T}|\\
    & = & \O(\Delta_X^3)\O(\mathbb{E}(|Span(I)|))\\
    & = & \O(\Delta_X^3)\O_{\td}(\Delta_X^4 \cdot |I|)\\
    & = & \O_{\td}(\Delta_X^7) \cdot |X|.
    \end{eqnarray*}
\end{proof}

\section{Negative results}
\label{sec:negative}

The next theorem follows directly from known bounds on the treedepth of planar graphs.

\begin{theorem}\label{thm:negativetwtd}
Under \ETH, there is no algorithm deciding an input $(G,k)$ of $\VC$ in time $n^{\O(1)}2^{o(\td(G))}$, even when restricting the input to planar graphs.
\end{theorem}
Note that this implies that under \ETH, there is no $n^{\O(1)}2^{o(\tw(G))}$-time algorithm neither.
\begin{proof}
The result for $\td$ boils down to observing (see for example~\cite{reidl2014faster}) that every planar graph $G$ satisfies $\td(G)=\O(\sqrt{n})$. Therefore, an algorithm for $\VC$ in time $n^{\O(1)}2^{o(\td(G))}$ on planar graphs would yield an algorithm in time $n^{\O(1)}2^{o(\sqrt{n})}$ for $\VC$ on planar graphs.
This would contradict $\ETH$~\cite[Lemma 5.1]{cai2003existence}.
\end{proof}

Although the next theorem does not seem to have been stated
explicitly in the literature, it follows immediately from a standard
reduction.

\begin{theorem}\label{thm:negativedeg3}
Unless $P = NP$, the problem $\VC/\disto{(\Delta(G)\leq3)}$, parameterized by the size of the modulator, is not in $XP$, even on planar graphs. In other words, there is no algorithm deciding an input $(G,X,k)$ of $\VC/\disto{(\Delta(G)\leq3)}$ in time $n^{f(|X|)}$, even when restricting the input to planar graphs.
\end{theorem}

\begin{proof}
It is known that $PLANAR$-3-$SAT$ is NP-complete~\cite[Theorem 2]{lichtenstein1982planar}. Moreover, the standard reduction from $PLANAR$-3-$SAT$ to \VC, establishing that \VC is NP-complete on planar graphs~\cite[Theorem 3]{lichtenstein1982planar}, produces graphs that are not only
planar but also subcubic. Hence, the resulting instances are already valid instances of $\VC/\disto{(\Delta(G)\leq 3)}$ with $X = \emptyset$.

Therefore, an algorithm for $\VC/\disto{(\Delta(G)\leq3)}$ in time $n^{f(|X|)}$ on planar graphs would yield a polynomial-time
algorithm for $PLANAR$-3-$SAT$. Since $PLANAR$-3-$SAT$ is NP-complete, this would imply
$P = NP$.
\end{proof}

\section{Conclusion and open problems}
\label{sec:conclusion}

\paragraph*{Open cases}
In this article, we provided subexponential algorithms for $\VC/\disto{\mathcal{C}}$ for several choices of $\mathcal{C}$ and input graph class $\mathcal{G}$.
Among the combinations of $\mathcal{C}$ and $\mathcal{G}$ that remain open, some particularly natural cases are the following.
(1) Does $\VC/\disto{(td=\O(1))}$ admit a subexponential FPT algorithm
on the class of intersection graphs of balls (resp. spheres) in $\mathbb{R}^d$? Note that these graphs are not TD-Mod compliant for $d\ge 3$, as the following construction shows that they cannot have modulator bounded treewidth. Let $X$ be a set of disjoint balls of radius $|X|$, then it is possible to add ${|X|}\choose{2}$ disjoint (long) paths made of radius one balls connecting any pair of balls from $X$. In that example $G\setminus X$ should have large treedepth. We conjecture that these graphs would have modulator bounded treewidth and bounded large neighborhood components, if these two notions would ask for the treedepth of $G\setminus X$ to be bounded. If this is the case, then our algorithm would work in the same way. 
(2) Does $\VC/\disto{\text{paths}}$ admit a subexponential FPT algorithm on string graphs ? Notice that our current algorithm requires $p$-bounded neighborhood complexity, which does not hold for string graphs as one can consider a clique $X$, together with $2^{|X|}$ extra vertices outside the clique such that for any $Z \subseteq X$, there is one vertex $v_Z$ such that $N_X(v_Z)=Z$.
Another natural case when $\mathcal{G}$ is a PATH-Mod compliant class (or even consider $s$-string graphs for example), is (3) whether $\VC/\disto{\mathcal{C}}$ admits a subexponential FPT algorithm when $\mathcal{C}$ is the class of forests ? Or bounded bridge-depth graphs ? Or even bounded treewidth ?

\paragraph*{Parameter reduction}
A notable aspect about this work concerns parameter reduction.
As already pointed out in \cite{10.1007/978-3-030-86838-3_1}, ordinary kernelization measures the "wrong" quantity (vertex size) when the eventual solver is exponential in a parameter (size of solution or $\tw$ typically) rather than in input size.
Indeed, given an instance $(G,k)$ of \VC, pipelining a kernel (to obtain an equivalent $(G',k')$) and solving $(G',k')$ using a dynamic programming by $\tw$ in time $f(\tw(G'))|V(G')|^{\O(1)}$ might not be significantly faster than solving directly $(G,k)$ in time $f(\tw(G))|V(G)|^{\O(1)}$.
However, as mentioned in \cite{10.1007/978-3-030-86838-3_1}, one cannot hope for a polynomial preprocessing algorithm that always output an equivalent instance with a strictly smaller parameter, nor an instance with  $\tw(G')=o(k)$.

A positive aspect of the framework considered in this paper (combining geometric graph classes and dist-to-triviality parameters) is that it provides a setting in which a provable $\tw$ reduction is possible.
Indeed, one can observe that all algorithms presented in this paper follow the same strategy: given $(G,X,k)$, they perform subexponential ($2^{o(|X|)}n^{\O(1)}$) branching to obtain graphs $G'$ such that $\tw(G')=o(|X|)$ (which are decided using classical dynamic programming by $\tw$).
Thus, this constitutes an example in which preprocessing can provably reduce a parameter that directly governs the running time. We can mention two results in this spirit. In the first one \cite{10.1007/978-3-030-86838-3_1}, authors consider the feedback vertex set problem, introduce the notion of antler, and prove that it can be decided if an antler of width $x$ exists (in FPT time by $x$),  which in this case allows to decrease the parameter $k$ by $x$.
In the second one \cite{marx2009treewidth}, authors provide an FPT time $\tw$ reduction that preserves relevant minimal separators.
Coming back to our results, this raises the two following questions. Can we reduce the treewidth to $o(|X|)$ (in geometrical classes) in polynomial time, hence corresponding to a polynomial kernel whose size is measured by $\tw$ ?
More generally, can similar preprocessing techniques provably reduce
parameters other than treewidth?

\paragraph*{Negative results}
The only negative result we have is \autoref{obs:lbASQGM} for $\mathcal{C}$ with unbounded treewidth, and even rules out XP algorithms. It would be very interesting to obtain negative results when $\tw(\mathcal{C})$ is bounded, and that precisely rule out subexponential algorithms. 
One difficulty in designing such reductions is the lack of suitable source problem. For instance, it appears to be open to determine if SAT parameterized by the number of vertices admits a subexponential algorithm when the clause-variable incidence graph is a string graph.
As a weaker form of negative result, ruling out the type of polynomial-time treewidth reduction discussed above could constitute an interesting intermediate objective.

\paragraph*{AI disclosure}
LLMs were used for the writing but not when conducting this research.

\bibliography{bib}

\end{document}